\documentstyle{mn}

\title[OH maser disc and outflow in Orion-BN/KL]
{OH maser disc and outflow in the Orion-BN/KL region}
\author[R. J. Cohen et al.]
{R. J. Cohen\thanks{Email:  rjc@jb.man.ac.uk}$^{1}$, 
N. Gasiprong,$^{1,2}$ 
J. Meaburn$^{1,3}$ and M. F. Graham$^{1}$.\\
$^{1}$University of Manchester, Jodrell Bank Observatory, 
Macclesfield, Cheshire, SK11 9DL \\
$^{2}$Department of Physics, University of Ubon Ratchithani, 
Ubon Ratchithani, Thailand \\ 
$^{3}$Instituto de Astronomia, UNAM, Apdo, Postal 877, Ensenada, BC 22800, 
Mexico}

\date{}
\begin{document}
\maketitle

\begin{abstract}
MERLIN measurements of 1.6-GHz OH masers 
associated with Orion-BN/KL are presented, and the data are compared with data
 on other masers, molecular lines, compact radio continuum sources 
and infrared sources in the region.  
OH masers are detected over an area 30~arcsec in diameter, with 
the majority lying along an approximately E-W structure that extends  
for $\sim$18~arcsec, encompassing the infrared sources IRc2, IRc6 and IRc7.  
Radial velocities range from --13 to +42~km~s$^{-1}$.  
The system of OH masers shows a velocity gradient together 
with non-circular motions.  The kinematics are modelled in terms of 
an expanding and rotating disc or torus. The rotation axis is found to be in 
the same direction as the molecular outflow.  There is an inner cavity of 
radius $\sim$1300~au with no OH masers.  The inner cavity, like the 
H$_{2}$O `shell' masers and SiO masers, is centred on radio source I. 
Some of the OH masers occur in velocity-coherent strings or 
arcs that are longer than 5~arcsec (2250~au).  One such feature, 
Stream A, is a linear structure at position angle $\sim$45\degr, lying between 
IRc2 and BN.  We suggest that these masers trace shock fronts, and have 
appeared, like a vapour trail, 200~yr after the passage of the runaway star 
BN.  The radio proper motions of BN, source I and source n project back to a 
region near the base of Stream A that is largely devoid of OH masers.   
The 1612-MHz masers are kinematically distinct from the other OH masers.  
They are also more 
widely distributed and appear to be associated with the outflow as traced by 
H$_{2}$O masers and by the 2.12-$\mu$m emission from shocked H$_{2}$.  
The magnetic field traced by the OH masers ranges from 1.8 to 16.3~mG, with 
a possible reversal.  
No OH masers were found associated with even the most prominent proplyds 
within 10~arcsec of $\theta^{1}$ Ori C.  
\end{abstract}



\section{Introduction}

The Orion A molecular cloud (L1641) is the nearest giant molecular cloud (GMC) 
containing the nearest regions of massive star-formation.  Because of its 
proximity it suffers little galactic extinction, and has been studied in far 
greater detail than other comparable regions in the Galaxy.    
The most recent  formation of high--mass stars in the Orion GMC is within the 
Becklin--Neugebauer--Kleinmann--Low 
(BN/KL) region, about 1~arcmin Northwest from  the Trapezium star $\theta^{1}$ 
Ori C.  The explosive nature of this event is demonstrated
dramatically by the infrared H$_{2}$ and [FeII] images of Allen
\& Burton (1993) and more recently Kaifu et al. (2000). `Bullets' of
matter are being ejected at hundreds of km s$^{-1}$ to form  
a bipolar cone of molecular `fingers' whose axis is 
perpendicular to the hot core 
traced by NH$_{3}$ emission (Wilson et al. 2000). These  `fingers' are
tipped by Herbig--Haro (HH) objects whose shocked gas is visible
at optical wavelengths (see Graham, Meaburn \& Redman 2003 for a recent 
association of these phenomena). 

At radio and millimetre wavelengths there are at least two molecular outflows, 
one associated with the optical and near-infrared features just described, and 
a second low-velocity outflow first detected through proper motions 
of H$_{2}$O masers (Genzel et al. 1981).  The powerful infrared source IRc2 
was originally thought to be the source of the outflow, but more recent 
observations, in particular of SiO masers (e.g. Greenhill et al. 1998), 
suggest that radio continuum source I of Menten \& Reid (1995) is the more 
likely outflow source.   Source I is significantly offset (by $\sim$0\farcs5) 
to the South of IRc2.  

Source I and the radio counterpart to BN have proper motions away from each 
other, corresponding to transverse speeds of 12 and 27~km~s$^{-1}$ 
respectively (Plambeck et al. 1995;  Rodr\'{i}guez et al. 2005).  Here and 
elsewhere we assume a distance of 450~pc.  The radio  proper motions indicate 
that BN and source I were within 225~au of each other 525$\pm$30~years ago.  
Tan (2004) has suggested that the large proper motion of BN is due to its 
ejection from the $\theta^{1}$~Ori~C system $\sim$4000~years ago.  Bally \& 
Zinnecker (2005) on the other hand propose that the motions of BN and I are 
the result of a stellar merger $\sim$500~years ago, which ejected BN and 
produced the energy powering the molecular outflow.  Further analysis of the 
radio proper motion data by G\'{o}mez et al. (in press) shows that a third 
source n also has proper motions that project back to the position on the sky 
where BN and source I were close, $\sim$500~years ago.  This raises the 
possibility that n is the third star dynamically neccessary to eject BN from 
the multiple system.  Distinguishing between these different possibilities is 
crucial to understanding the history of recent star-formation in the region.  

We have observed hydroxyl (OH) masers in the Orion-BN/KL region as part of an 
ongoing study of masers associated with molecular outflows from massive young 
stars (Hutawarkorn \& Cohen 2005, and references therein).  OH masers have 
the potential to trace physical conditions, including magnetic fields which 
are detected and measured through Zeeman splitting and OH maser polarization.  
Previous MERLIN OH observations by Norris (1984) detected 80 1665-MHz 
masers distributed over a region of 10~arcsec, which were interpreted in 
terms of a rotating torus surrounding IRc2.  
Johnston, Migenes \& Norris (1989) detected 175 masers spread over a region of 
20~arcsec, with the majority in an E-W structure 14~arcsec in extent.  
The masers were found in clusters or  `clumps'  that correlated with NH$_{3}$ 
emission.  
The present MERLIN observations more than double the number of OH 
masers known and show that the distribution is even more extensive in 
position and velocity (Section~\ref{distribution}).  The association of the 
masers with radio continuum sources is described in Sections~\ref{distribution}
 and~\ref{arcs}, while the association with near- and mid-infrared sources is
 described in Section~\ref{comparison}.   

Secondary targets in the same MERLIN observations were the `proplyds' (O'Dell,
Wen \& Hu 1993) 
in the vicinity of $\theta^{1}$ Ori C. These are the solar system--sized
circumstellar envelopes discovered 
by Laques \& Vidal (1979) which surround low--mass YSOs (Churchwell et al
1987; Meaburn 1988; McCaughrean \& Stauffer 1994; O'Dell \& Wong 1996;
Bally et al. 1998a). These stellar cocoons have dusty molecular disks with
dense ($\geq$ 10$^{6}$ cm$^{-3}$) surfaces ionized by the intense 
flux of Lyman radiation from $\theta^{1}$ Ori C (see Graham et al. 2002). 
In these circumstances
maser emission could be anticipated from the molecular disks which, 
if detectable, could be an important tool for investigating proto-planetary
environments.  Upper limits for such emission are reported in 
Section~\ref{proplyds1}.

\section{Observations}

The OH observations were performed 
 on 1998 April 27 and 28 using seven telescopes of MERLIN 
(the Multi Element Radio Linked Interferometer Network):  the
76-m Lovell Telescope and the Mk2 telescope at Jodrell Bank, and 
outstation telescopes at Pickmere, Darnhall, Knockin, Defford and Cambridge.  
The longest baseline was 218~km, giving a minimum fringe spacing of 
0.16~arcsec.  Observations covered all four 
ground-state OH transitions, namely 1612, 1665, 1667 and 1720~MHz.  A spectral
 bandwidth of 500~kHz was used, corresponding to 
a velocity range of 90 km~s$^{-1}$.  
The spectral band was divided into 
256 frequency channels, giving a velocity spacing of 
0.35 km~s$^{-1}$.  
The radial velocity at the centre of the 500-kHz band was +28~km~s$^{-1}$ 
with respect to the local standard of rest (lsr).  
Left-left and right-right circularly
polarized signals from each pair of telescopes were simultaneously
correlated to give LL and RR.  

The OH lines were observed in pairs, 
1612 and 1667~MHz for 6.5 hours on 27 April 1998 and  
1665 and 1720~MHz for 8 hours on 28 April 1998. 
Observations consisted of 
5-min scans tracking the field centre 
RA(B1950) 05$^{\rm h}$32$^{\rm m}$49\fs043, 
Dec. (B1950) --05\degr25\arcmin16\farcs003  
at each OH line frequency interleaved with 4-min
scans on the nearby unresolved phase calibrator source IC0539-057.  
Because of the faintness of the phase calibrator source it was 
necessary to observe it in 16-MHz bandwidth.  
Short ${\sim}$ 30-min scans of the point source amplitude and bandpass 
calibrator 2134+004 were also made.  2134+004 was also used to 
provide the flux scale (based on comparisons with 3C286).

The data processing and analysis procedures were carried out in 
B1950 coordinates, as described by 
Gasiprong, Cohen and Hutawarakorn (2002).  
Data were first edited, calibrated 
 and corrected for gain-elevation effects using the 
Jodrell Bank d-programmes, and then passed into the AIPS software package.  
Within the AIPS package the data were further calibrated for all remaining 
instrumental and atmospheric effects including the instrumental
polarization.
Self-calibration imaging of a
bright pointlike reference channel was employed to derive
the final gain solutions for the 1665-MHz data.  
These gain solutions were then applied to
the other channels.  The 1612- and 1667-MHz data sets had poorer 
signal-to-noise and could not be self-calibrated, while at 
1720~MHz no emission was detected at all.  
Finally, maps in each circular polarization 
(RHC and LHC) 
were produced using CLEAN algorithms in AIPS, with a 
restoring beam of 150~mas~$\times$~150~mas.  For each data cube 
the pixel size was 45$\times$45~mas$^{2}$ and the total area mapped 
was 1024$\times$1024 pixels$^{2}$ (46$\times$46~arcsec$^{2}$), centred 
on the position of IRc2.  A second field of 512$\times$512 pixels 
 (23$\times$23~arcsec$^{2}$) 
centred on the position of  $\theta^{1}$ Ori C was also searched 
for possible emission associated with the proplyds.  

The rms-noise in the final maps was 30-50~mJy~beam$^{-1}$, so only 
masers brighter than  $\sim$0.1~Jy could be detected.  
The masers were all unresolved at 150-mas resolution.  
Positions of maser 
features were determined by fitting two-dimensional Gaussian 
components to the brightest peaks in each channel map and taking 
flux-weighted 
means across the channels showing emission from each particular feature.  
Final B1950 coordinates of masers were then converted to J2000 using the 
software package coco.  The errors in the relative positions are typically 
10~mas, while the absolute positions have a possible systematic error of 
20~mas due to phase-transfer from the reference source.  

\section{Results}

\subsection{Maser distribution in Orion-BN/KL}
\label{distribution}

\begin{figure*}
\vspace{225mm}
\includegraphics{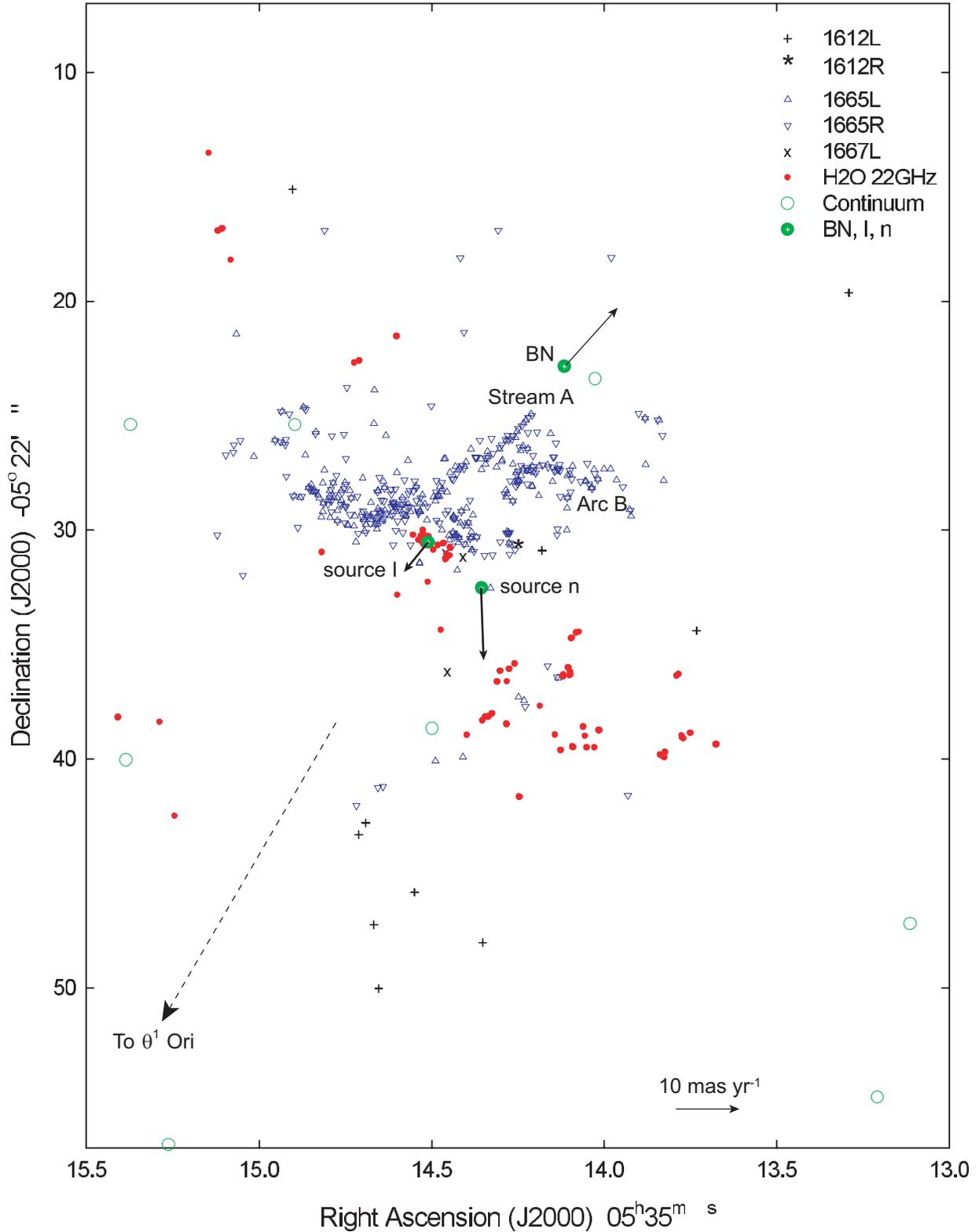}
\caption{Positions of OH masers (blue symbols, this paper),  H$_{2}$O masers 
(red symbols, Gaume et al. 1998), and compact continuum sources (green circles,
 Zapata et al. 2004) in Orion-BN/KL.  The central region around IRc2 and radio 
source I is shown on an expanded scale in Fig.~\ref{ra-deccore}.  The radio 
continuum sources BN, I and n are highlighted as filled green circles, and 
their proper motions are indicated schematically (data from Rodr\'{i}guez et 
al. 2005 and G\'{o}mez et al. 2005).  The direction to  $\theta^{1}$ Ori C is 
indicated by the dashed arrow.  }
\label{ra-dec}
\end{figure*}

\begin{figure}
\vspace{112mm}
\includegraphics{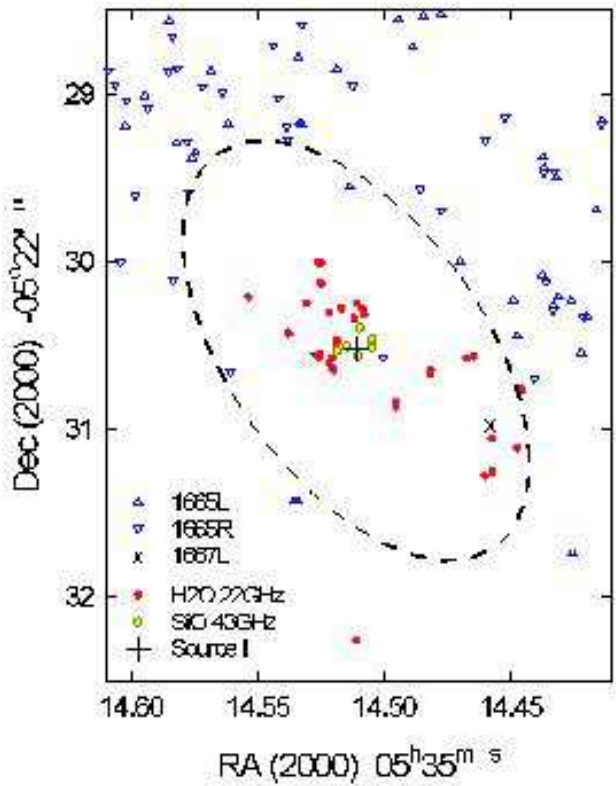}
\caption{Positions of OH masers (blue symbols, this paper), H$_{2}$O masers 
(red symbols, Gaume et al. 1998) and SiO maser clusters (yellow symbols, 
Doleman et al. 1999) in the central region of Orion-KL, near radio source I 
(Menten \& Reid (1995).  The position for source I corresponds to epoch 2000 
(Rodr\'{i}guez et al. 2000).  The dashed ellipse corresponds to a ring  
with rotation axis at position angle --35\degr and inclination angle 58\degr 
to the line-of-sight, centred on radio source I (see Section~\ref{model}).}
\label{ra-deccore}
\end{figure}

We detected 11 masers at 1612~MHz, 430 masers at 1665~MHz, 3 masers at 
1667~MHz (LHC only) and no masers at 1720~MHz.  The positions and velocities 
are given in Table~\ref{OHtab}.   
To facilitate comparison with other data, 
we give both the B1950 coordinates and the J2000 coordinates.  
The maser distribution is shown in Fig.~\ref{ra-dec} in J2000 coordinates.  
The OH masers are spread over a region 30~arcsec
in extent, corresponding to $\sim$13500~au.  
The radial velocities range from --13 to +42~km~s$^{-1}$.  
The distribution is more extensive in both angular scale and velocity 
range than found in previous work.  However most of the emission comes from 
the dominant E-W structure noted by Norris (1984) and Johnston et al. (1989).

One striking aspect of the new maps is that 
many of the OH masers lie in semi-continuous streams or arcs, 
of which the most prominent are labelled Stream A and Arc B in 
Fig.~\ref{ra-dec}.  
Stream A stretches from RA (2000) 05$^{\rm h}$35$^{\rm m}$14\fs2, 
Dec. (2000) --05\degr22\arcmin25\arcsec to 05$^{\rm h}$35$^{\rm m}$14\fs5, 
--05\degr22\arcmin28\arcsec, and includes the maser `clump' NW1 identified by 
Johnston et al. (1989).  The radial velocities are approximately constant, 
around +21$\pm$2~km~s$^{-1}$.  Arc B stretches 
from 05$^{\rm h}$35$^{\rm m}$14\fs1,\linebreak[4] --05\degr22\arcmin28\arcsec 
to 05$^{\rm h}$35$^{\rm m}$14\fs3, --05\degr22\arcmin28\arcsec, and includes 
the 
maser clumps NW2 and NW3 of Johnston et al. (1989).   The radial velocities 
cover a wide range, from $\sim$-7~km~s$^{-1}$ in the East to  
$\sim$+10~km~s$^{-1}$ in the West. 
These streams or arcs are large-scale features, with angular sizes of 
$\sim$5~arcsec that 
correspond to $\sim$2250~au.  We argue that they 
trace large-scale shocks or ionization fronts (Section~\ref{arcs}).  

The distribution of H$_{2}$O masers from Gaume et al. (1998) is also shown 
in  Fig.~\ref{ra-dec}.  The OH and H$_{2}$O masers 
have complementary distributions, with almost no overlap.  
An expanded plot of the crowded 
central region is shown in Fig.~\ref{ra-deccore}, revealing the same 
situation on the small scale.   The  H$_{2}$O masers in this region are the 
so-called `shell' masers that were not detected in the VLBI measurements by 
Genzel et al. (1981).  There is an elliptical zone of avoidance centred on 
source I, within which there are very few OH masers, but the H$_{2}$O `shell' 
masers and SiO masers are found.  
The  H$_{2}$O `shell'  masers lie inside the dashed ellipse in the figure, 
while nearly all of the OH masers lies outside the ellipse.  The position
 angle and inclination angle of the ellipse were chosen to match the model 
fitted to the OH masers in Section~\ref{model}.  

Fig.~\ref{ra-deccore} also shows the positions of the SiO maser clusters 
mapped by Doeleman, Lonsdale \& Pelkey (1999).  The SiO masers cluster tightly 
around the position of the radio source I (Menten \& Reid 1995), with the 
H$_{2}$O `shell' masers surrounding them.
The relative location of SiO, H$_{2}$O shell and OH masers at 
increasing distances from the central source I is similar to that found 
for these masers in the circumstellar envelopes of post-AGB stars, 
which has a natural explanation in terms of excitation by a powerful central 
source  (Chapman \& Cohen 1986).

The positions of compact radio continuum sources in the region are plotted in 
Fig.~\ref{ra-dec} as open green circles (data from Zapata et al. 2004), with 
sources I, n and BN highlighted as filled green circles.  Apart from source I, 
which is at the centre of the SiO and H$_{2}$O `shell' maser distributions and 
the OH zone of avoidance, there is another association of OH masers with a 
compact radio continuum source that is noteworthy, namely   
a striking association between Stream A and BN.  The stream points directly 
towards BN (away from IRc2), in the direction of BN's proper motion (shown 
schematically by the arrow in Fig.~\ref{ra-dec}).  This is discussed further 
in Section.~\ref{arcs}.  
Finally we note a possible weaker association of source D of Menten \& Reid 
(1995) (source 21 of Zapata et al.), near  
RA (J2000)~=~05$^{\rm h}$35$^{\rm m}$14\fs90, Dec. (J2000)~= 
\linebreak--05\degr22\arcmin25\farcs4, with a cluster of OH masers, 
corresponding to `clump NE1' of Johnston et al. (1989).  The radial 
velocities are centred near +8~km~s$^{-1}$.  

Further study of Fig.~\ref{ra-dec} reveals that the OH 1612-MHz masers
are distributed differently from the mainline 1665- and 1667-MHz masers.  
Most of the OH 1612-MHz masers are found a long way from the centre, in the 
outskirts of the OH maser region, and they are spatially separated from the 
other OH masers.  
Their far-flung distribution is strikingly different from that of
the other OH masers, but similar to that of the H$_{2}$O masers.  
This similarity is reinforced by kinematic similarities.

\subsection{Maser kinematics}
\label{kinematics}

\begin{figure*}
\vspace{104mm}
\includegraphics{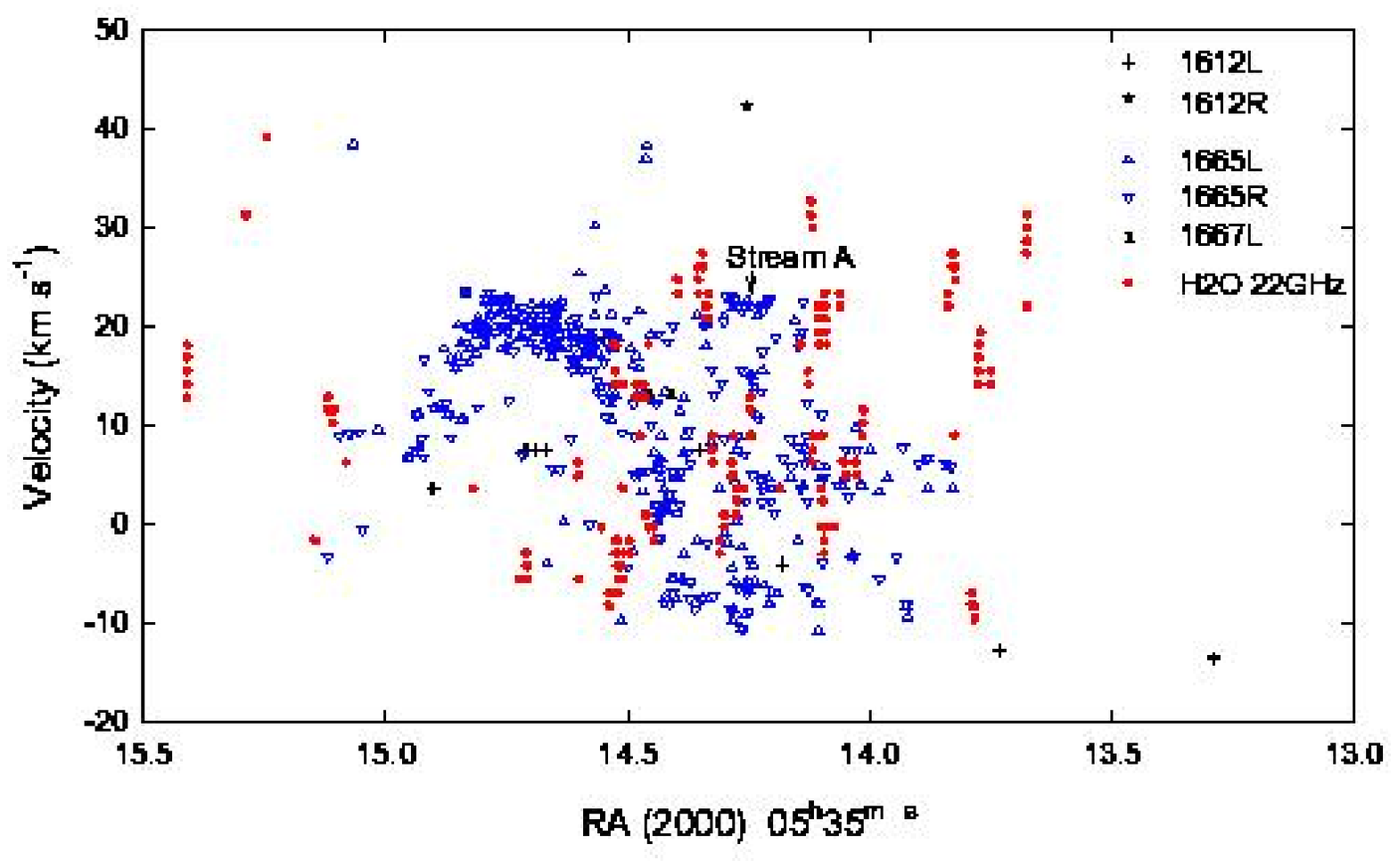}
\caption{Position-velocity plot showing right ascension (J2000) and  
radial velocity for the OH masers (blue symbols, this paper) and the 
H$_{2}$O masers (red symbols, Gaume et al. 1998) in Orion-BN/KL.  Stream A 
has a radial velocity of +21~km~s$^{-1}$ that is distinct from the general 
rotational pattern of the other OH masers.  
}
\label{ra-vel}
\end{figure*}

\begin{figure*}
\vspace{104mm}
\includegraphics{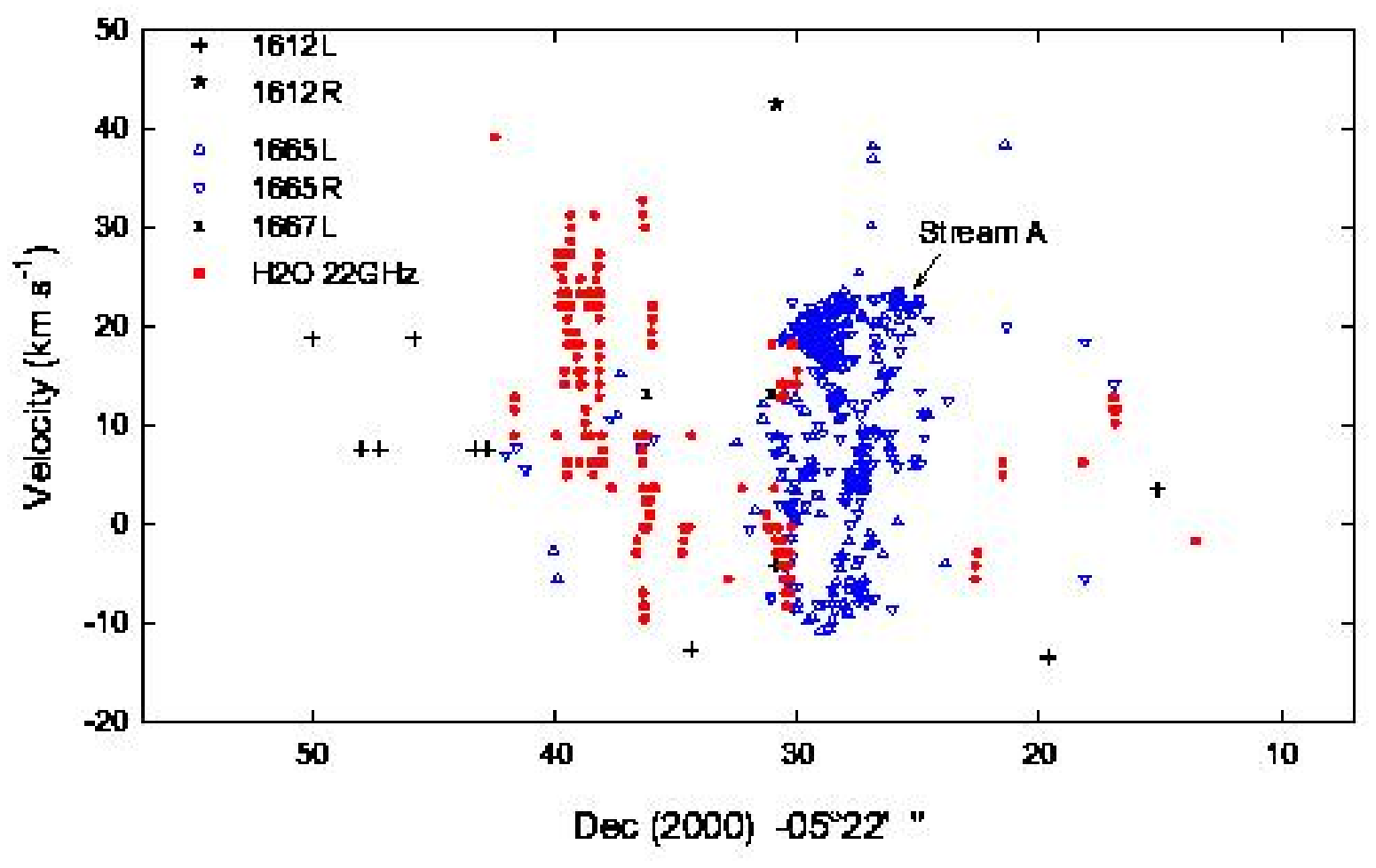}
\caption{Position-velocity plot showing declination (J2000) and radial 
velocity for the OH masers (blue symbols, this paper) and the H$_{2}$O 
masers (red symbols, Gaume et al. 1998) in Orion-BN/KL.  
}
\label{dec-vel}
\end{figure*}

The complex kinematics of the Orion-BN/KL masers are summarized in the position-velocity diagrams Figs.~\ref{ra-vel} and \ref{dec-vel}.  Fig.~\ref{ra-vel} shows RA vs. $V_{\rm lsr}$ for the OH and H$_{2}$O masers.  There is a clear contrast between the two species:  the OH masers show a dominant rotational pattern, in which the radial velocities drop from +20~km~s$^{-1}$ in the East to 0~km~s$^{-1}$ in the West, while the 
H$_{2}$O masers show a dominant expansion pattern, with highly red-shifted or blue-shifted masers equally likely to occur to the East or West.  This is consistent with the well known expansional proper motions of these masers (Genzel et al. 1981;  Gaume et al. 1998).  The OH rotational pattern is clearest at right ascensions greater than  05$^{\rm h}$35$^{\rm m}$14\fs5, and more disturbed at right ascensions less than this.  Stream A has a velocity greatly different from the rotational pattern, as indicated in  Fig.~\ref{ra-vel}.  We note that the radial velocity of Stream A, +21~km~s$^{-1}$, is the same as that of BN and its circumstellar nebulae within 25~au (Scoville et al. 1983).

Fig.~\ref{dec-vel} shows a Dec-$V_{\rm lsr}$ plot.  The OH masers trace a dominant inclined elliptical pattern, centred on Dec. (2000) --05\degr22\arcmin29\arcsec and $V_{\rm lsr}$~=~+6~km~s$^{-1}$.  Radial velocities rise abruptly from 0--10~km~s$^{-1}$ in the South to +20~km~s$^{-1}$ to the North of Dec. (2000) --05\degr22\arcmin29\arcsec.  The `hole' in the centre of the  Dec-$V_{\rm lsr}$ ellipse suggests that in addition to rotation there is an expansional component of motion of similar magnitude.  The locations of OH and H$_{2}$O masers in this diagram are again complementary.    

The OH 1612-MHz masers are kinematically and spatially distinct from the 1665- and 1667-MHz masers.  This is particularly clear in Fig.~\ref{dec-vel}, where they are well separated in both declination and velocity, appearing at the extremes of both coordinates, well separated from the rotating and expanding ring signature of the mainline masers.  The kinematic pattern is similar to but even more extensive than that of the H$_{2}$O masers, suggesting that outflow rather than rotation is the dominant motion.

The OH distribution and kinematics are modelled in Section ~\ref{model}.

\subsection{Comparison with Infrared Data}
\label{comparison}

The distribution and kinematics of the dominant OH emission show a degree of symmetry about the position of IRc2 and radio source I, and a radial velocity of $\sim$8~km~s$^{-1}$, as noted by previous authors cited in the introduction.  The OH masers encompass IRc2, 3, 6 and 7.  The availability of high resolution infrared data allows us to examine in more detail the correspondences of the OH masers with the powerful mid-infrared sources and with near-infrared bullets in the molecular outflow.

\subsubsection{Comparison with 12.5-$\mu$m Keck and 11.7-$\mu$m Gemini Data}
\label{midIR}

\begin{figure*}
\begin{center}
\vspace{140mm}
\includegraphics{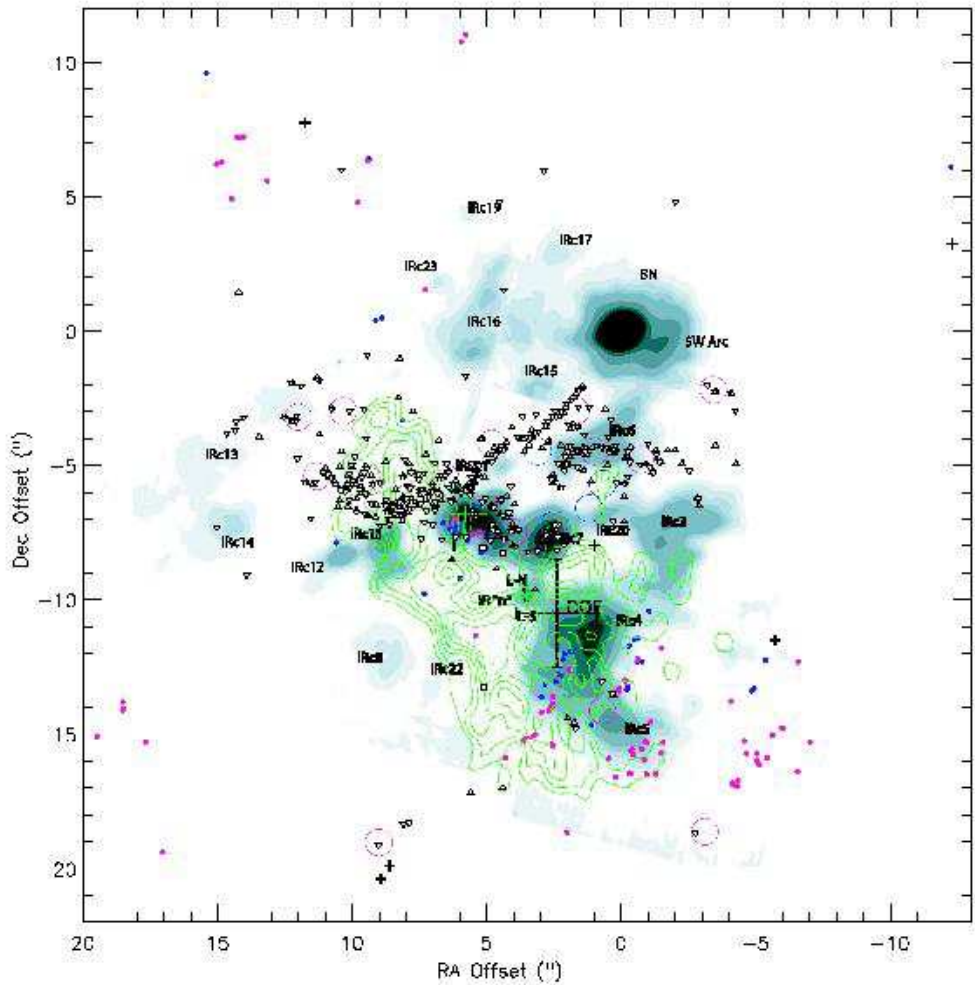}
\caption{Positions of OH masers (black symbols as in Fig.~\ref{ra-dec}) overlaid on the 
12.5-$\mu$m image (greyscale) taken with the Keck Telescope by Shuping et al. (2004), adapted from their Figure 5.  The open circles show the positions of OH maser clumps identified by Johnston et al. (1989).  Red and blue dots indicate H$_{2}$O masers from Gaume et al. (1998) and Genzel et al. (1981), with red indicating radial velocity greater than +5~km~s$^{-1}$ and blue indicating radial velocity less than +5~km~s$^{-1}$.  The green contours show NH$_{3}$ (J,K)~=~(4,4) emission associated with the `hot core', from Wilson et al. (2000).  Offsets are relative to the position of the BN object:  RA (J2000)~=~05$^{\rm h}$35$^{\rm m}$14\fs117, Dec. (J2000)~=~--05\degr22\arcmin22\farcs9.  IRc2 is near (+5\arcsec, --7\arcsec) and IRc6 is near (+1\arcsec,--4\arcsec).}
\end{center}
\label{keck}
\end{figure*}

Shuping, Morris \& Bally (2004) have published a high resolution (0\farcs38) mid-infrared image of the Orion BN/KL region obtained with the Keck I telescope at 12.5-$\mu$m.  
Fig.~5 is a composite plot showing the OH masers (this paper) superimposed on Fig.~5 of Shuping et al. (2004).  This shows several surprising results.  
Firstly, OH masers cluster around IRc2, but they avoid the actual source position (relative coordinates (+5\arcsec, --7\arcsec) in this diagram), which lies in the zone of avoidance noted in Section~\ref{distribution}.  
Secondly, Stream A points directly away from IRc2 towards the dominant mid-infrared source BN (at the origin), at position angle $\sim$45\degr.  Stream A runs through a region devoid of mid-infrared emission.  The direction from IRc2 to BN is parallel to the proper motions recently measured for BN and radio source I by Rodr\'{i}guez et al. (2005).  It is also the projected direction of the large-scale molecular outflow.  
Thirdly, Arc B curves parabolically around the extended source IRc6 (near (+1\arcsec,--4\arcsec) in this figure), in a manner that suggests a close physical association.  This confirms and extends the result of Johnston et al. (1989).

Smith et al. (2005) have published a high-resolution  11.7-$\mu$m mosaic image of the inner Orion nebula obtained with the Gemini South telescope.  
The only close correlation (better than 2 arcsec) of any of the OH 1.6-GHz
masers listed in Table~\ref{OHtab} with the 11.7-$\mu$m point sources listed by Smith et al. (2005) is the 1612L maser number 10, at 
 RA (J2000)~=~05$^{\rm h}$35$^{\rm m}$14\fs6533, Dec. (J2000)~=~-05\degr22\arcmin50\farcs022, which has a radial velocity of +18.9~km~s$^{-1}$.  The corresponding IR source is at 
05$^{\rm h}$35$^{\rm m}$14\fs67, --05\degr22\arcmin49\farcs5.  
Smith et al. (2005) describe the latter as `an embedded IR source
with no optical ID'. None of the 27 known proplyds that correlate with
11.7-$\mu$m point sources in Smith et al. (2005) are identified here as
maser sources.

\subsubsection{Comparison with 2.12-$\mu$m Subaru Data}
\label{nearIR}

\begin{figure*}
\begin{center}
\vspace{155mm}
\includegraphics{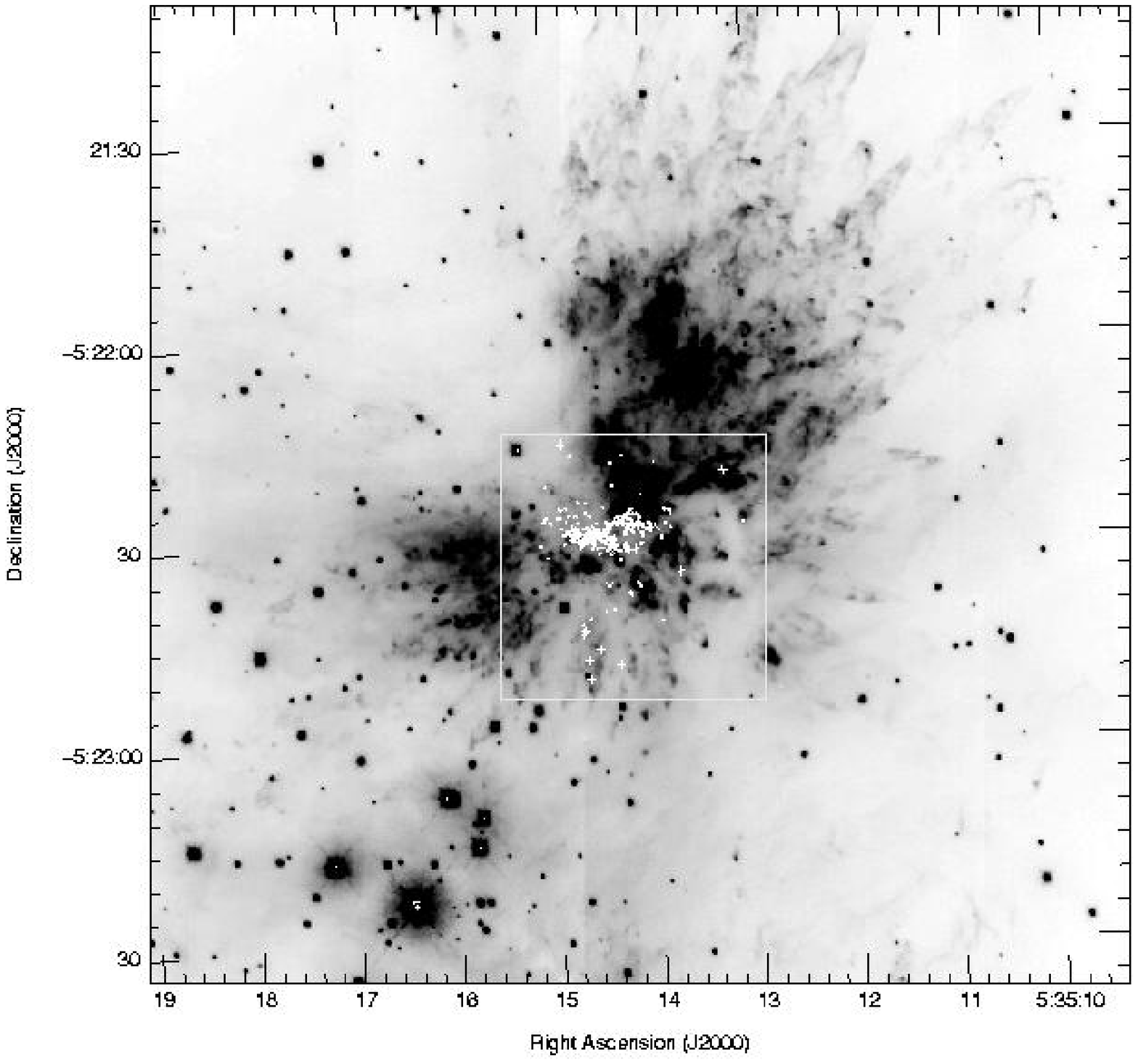}
\caption{Positions of the OH masers in Orion-BN/KL, overlaid on the infrared 
2.12-$\mu$m H$_{2}$ image taken with the Subaru Telescope by Kaifu 
et al. (2000), showing the relationship of the masers to the 
outflow (Section~\ref{comparison}).  The 1665-MHz masers (.) and the 1667-MHz masers (x) 
lie in an expanding and rotating disc or torus centred on IRc2 and source I (Section~\ref{model}), while the 1612-MHz masers (+) have a more widespread distribution.    
The area in the box is shown on an expanded scale in Fig.~\ref{subcore}.}
\label{subaru}
\end{center}
\end{figure*}

The locations of the OH masers relative to the well-studied molecular 
outflow are shown in Fig.~\ref{subaru}.  
The infrared H$_{2}$~2.12-${\rm \mu m}$ images presented here were taken
with the Subaru Telescope as part of its astronomical first light programme 
(Kaifu et al. 2000). The data
were taken on 1999 January 11, 13 and March 6 using the Cooled Infrared
Spectrograph and Camera for the OH-suppression spectrograph, CISCO 
(Motohara et al. 1998).

The 2.12-$\mu$m image shown in Fig.~\ref{subaru} is a 
150\arcsec $\times$ 150\arcsec section of the original 
$5\arcmin\times5\arcmin$ mosaic (provided by Masa Hayashi).  
In order to overlay the OH maser positions onto 
the Subaru image, it was first necessary to astrometrically register the 
image with the J2000 coordinate system.  This was done using \(>600\) stars 
from the catalogue of Hillenbrand \& Carpenter (2000) and the Starlink 
 {\sc gaia} software package.   
A computer programme was then written to read the B1950
OH maser positions and 
convert them to J2000 coordinates using the Starlink Astrometry Library
and produce a catalogue that was readable by the {\sc gaia} software
package. The positions were then overlaid using {\sc gaia}.

\begin{figure}
\begin{center}
\vspace{78mm}
\includegraphics{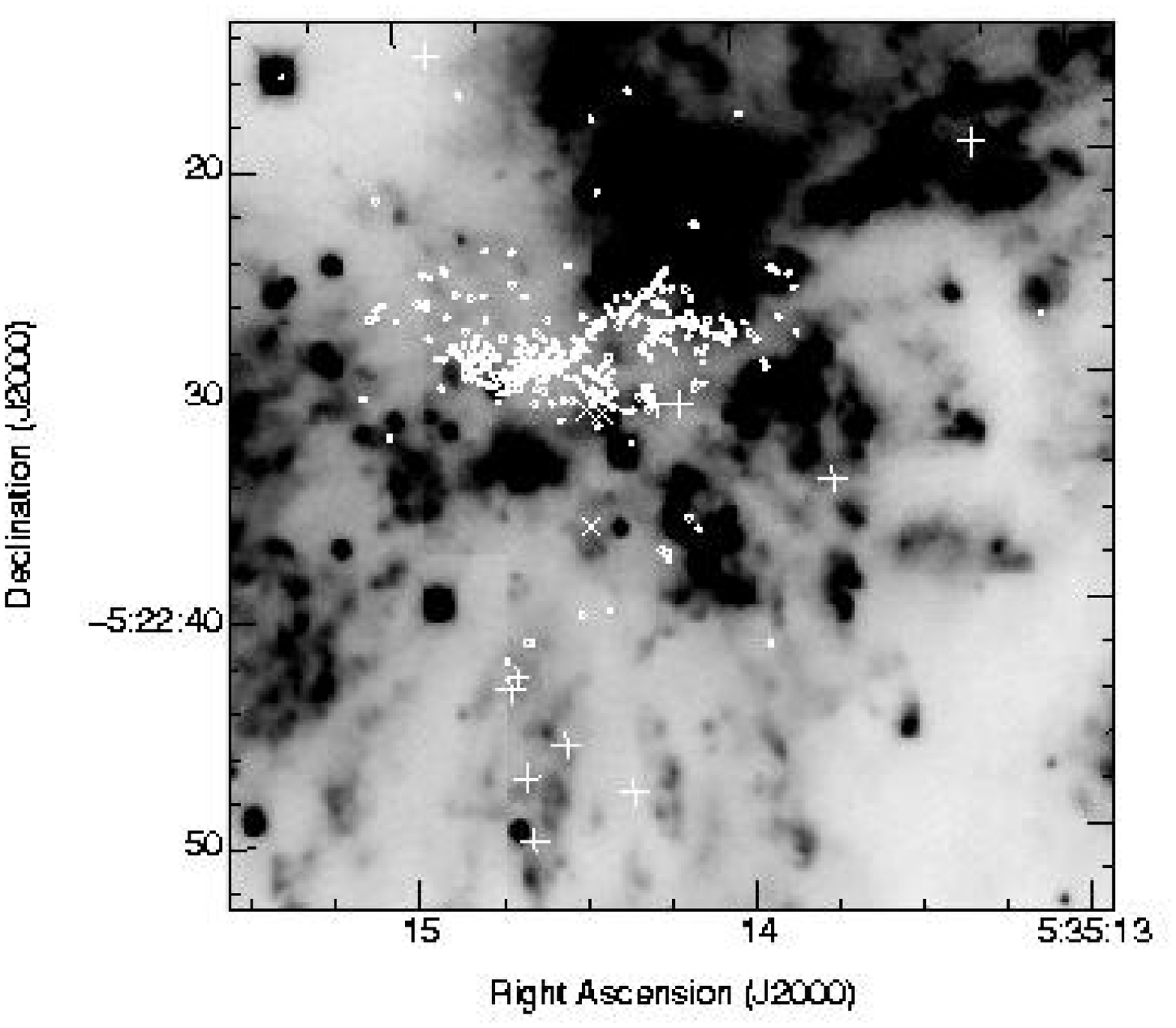}
\caption{Expanded view of the central region of Fig.~\ref{subaru}, showing 
the positions of OH masers overlaid on the infrared 
2.12-$\mu$m H$_{2}$ image taken with the Subaru Telescope by Kaifu 
et al. (2000).   
The 1612-MHz masers (+) are associated with fingers of 
H$_{2}$ emission in the molecular outflow.}
\label{subcore}
\end{center}
\end{figure}

The 1665- and 1667-MHz masers (shown by dot and cross symbols respectively) 
are concentrated towards the highly obscured IRc2 region.  In general they 
show little 
correspondence with features in the near-IR map.  However the 1612-MHz masers 
(+) correspond with fingers of H$_{2}$ emission in the molecular 
outflow.  This association is seen more clearly in Fig.~\ref{subcore}, 
which shows the central region on an expanded scale.  
The 1612-MHz masers  show a strong association with fingers of H$_{2}$ 
emission in the molecular outflow, particularly in the South, 
near RA (J2000) 05$^{\rm h}$35$^{\rm m}$14\fs7, 
Dec. (J2000) --05\degr22\arcmin46\arcsec, 
where 6 (half) of the 1612-MHz masers are found.  Three 1665R masers (numbers 
63, 64 and 76 in Table~\ref{OHtab}) are found nearby, and may also be 
associated.  
This is the first time that clear counterparts to interstellar OH masers have 
been seen.  It is not clear, however, why these particular H$_{2}$ emission 
fingers should have OH counterparts, while many others do not.

\subsection{Zeeman Splitting and Magnetic Fields}
\label{zeeman}

\setcounter{table}{1}
\begin{table*}
\caption{Possible OH Zeeman pairs in Orion.}
\label{zeemantab}

\begin{tabular}{lcccccccc}
\hline
Zeeman & Transition & V$_{\rm lsr}$ & RA (1950) & Dec. (1950) &  RA (2000) & 
Dec. (2000) & S$_{\rm pk}$ & B \\
Pair   &    & km~s$^{-1}$   &  05$^{\rm h}$32$^{\rm m}$  $^{\rm s}$ 
& --5$^{\circ}$:24\arcmin:~~\arcsec &  05$^{\rm h}$35$^{\rm m}$  
$^{\rm s}$ & --5$^{\circ}$:22\arcmin:~~\arcsec &(Jy)  & mG \\
\hline  
Z$_{1}$ & 1665L & 21.98  & 46.8077  & 19.522 & 14.2830 & 26.065 & 0.18 & 
+1.8 \\  
        & 1665R & 23.06  & 46.8102  & 19.504 & 14.2855 & 26.047 & 0.10 & 
     \\
        &       &        &          &        &      &      \\
Z$_{2}$ & 1665L & 18.13  & 47.3239  & 21.992 & 14.7984 & 28.572 & 0.10 & +3.6
 \\
        & 1665R & 20.24  & 47.3210  & 22.020 & 14.7954 & 28.600 & 0.38 &    
  \\
        &       &        &          &        &      &      \\
Z$_{3}$ & 1665L & 15.67  & 46.7458  & 19.896 & 14.2210 & 26.434 & 0.14 & --3.5 
\\
        & 1665R & 13.56  & 46.7487  & 19.915 & 14.2239 & 26.454 & 0.22 &      
\\
        &       &        &          &        &      &      \\
Z$_{4}$ & 1665L & 11.59  & 46.9193  & 23.131 & 14.3934 & 29.682 & 0.26 & +16.3 
\\
        & 1665R & 1.96   & 46.9203  & 23.109 & 14.3944 & 29.660 & 0.10 &      
\\
        &       &        &          &        &      &      \\
Z$_{5}$ & 1665L & 0.30   & 46.8021  & 23.549 & 14.2761 & 30.092 & 0.58 & +3.6 
\\
        & 1665R & 2.43   & 46.7989  & 23.565 & 14.2729 & 30.108 & 1.48 &      
\\
\hline
\end{tabular}
\end{table*}

The data were searched for possible Zeeman pairs of opposite circular 
polarization, with radial velocities differing by more than 0.8~km~s$^{-1}$ 
(to exclude linearly polarized features) and lying within 30~mas of each 
other.  The results are given in Table~\ref{zeemantab}.    We estimate the 
chance of a false association to be no more than 4~percent.  

\begin{figure}
\begin{center}
\vspace{84mm}
\includegraphics{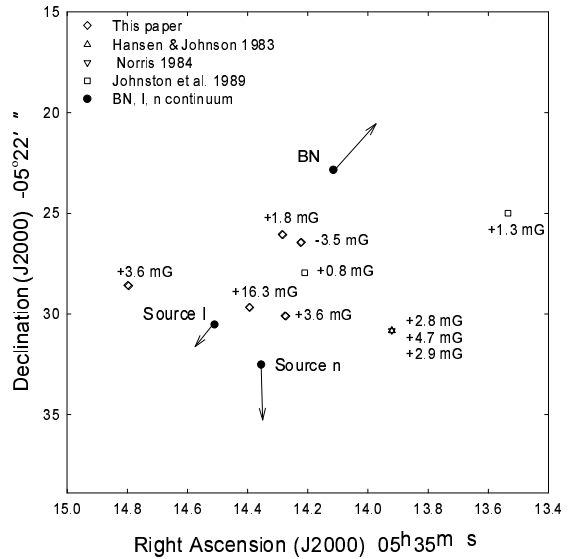}
\caption{Magnetic field measurements in the Orion-BN/KL region from OH maser 
Zeeman splitting (present paper, plus references indicated in the legend).  
The locations and schematic proper motions of the continuum sources BN, I 
and n are shown for reference, as in Fig.~\ref{ra-dec}.} 
\label{magnetic}
\end{center}
\end{figure}

The magnetic fields range from --3.5~mG (directed towards us) to +16.3~mG.  
In Fig.~\ref{magnetic} we plot the magnetic field values as a function of 
position on the sky, together with values from 
previous interferometric OH Zeeman measurements (Hansen \& Johnston 1983; 
Norris 1984;  Johnston et al. 1989).  The present data suggest that in 
addition to the overall field of 1--3~mG directed away from us, there may 
also be regions of higher field strength (Z$_{4}$), and that the field 
direction may reverse (Z$_{3}$).  The field reversal is seen at only a 
single point, and not on the larger-scale that is found in other bipolar 
outflow sources (Blaskiewicz et al. 2005; Hutawarakorn \& Cohen 2005, and 
references therein).  More data are needed to confirm these findings.

\subsection{Search of the Proplyd Region}
\label{proplyds1}

A search of a 23~arcsec$\times$23~arcsec region around $\theta^{1}$ Ori C was 
made, but no OH masers with peak flux densities $\geq$0.1~Jy were found, even 
near the 
most prominent proplyds (LV 1--6).  
The densities in these regions have been estimated to be 
10$^{5}$--10$^{6}$~cm$^{-3}$ (Section~\ref{proplyds2}), which are typical of 
OH maser sources, while the temperatures are estimated to be 200--350~K 
(Hayward, Houck \& Miles 1995), which are somewhat higher than usually 
encountered in OH maser sources (e.g. Cragg, Sobolev \& Godfrey 2002).  
The non-detection of OH masers might be explained by a low OH column density, 
which could arise through chemical evolution  (e.g. Rodgers \& Charnely 2001).

\section{Discussion}
\label{discussion}

\subsection{Kinematic Model for OH 1665-MHz Masers}
\label{model}

\begin{figure*}
\begin{center}
\vspace{155mm}
\includegraphics{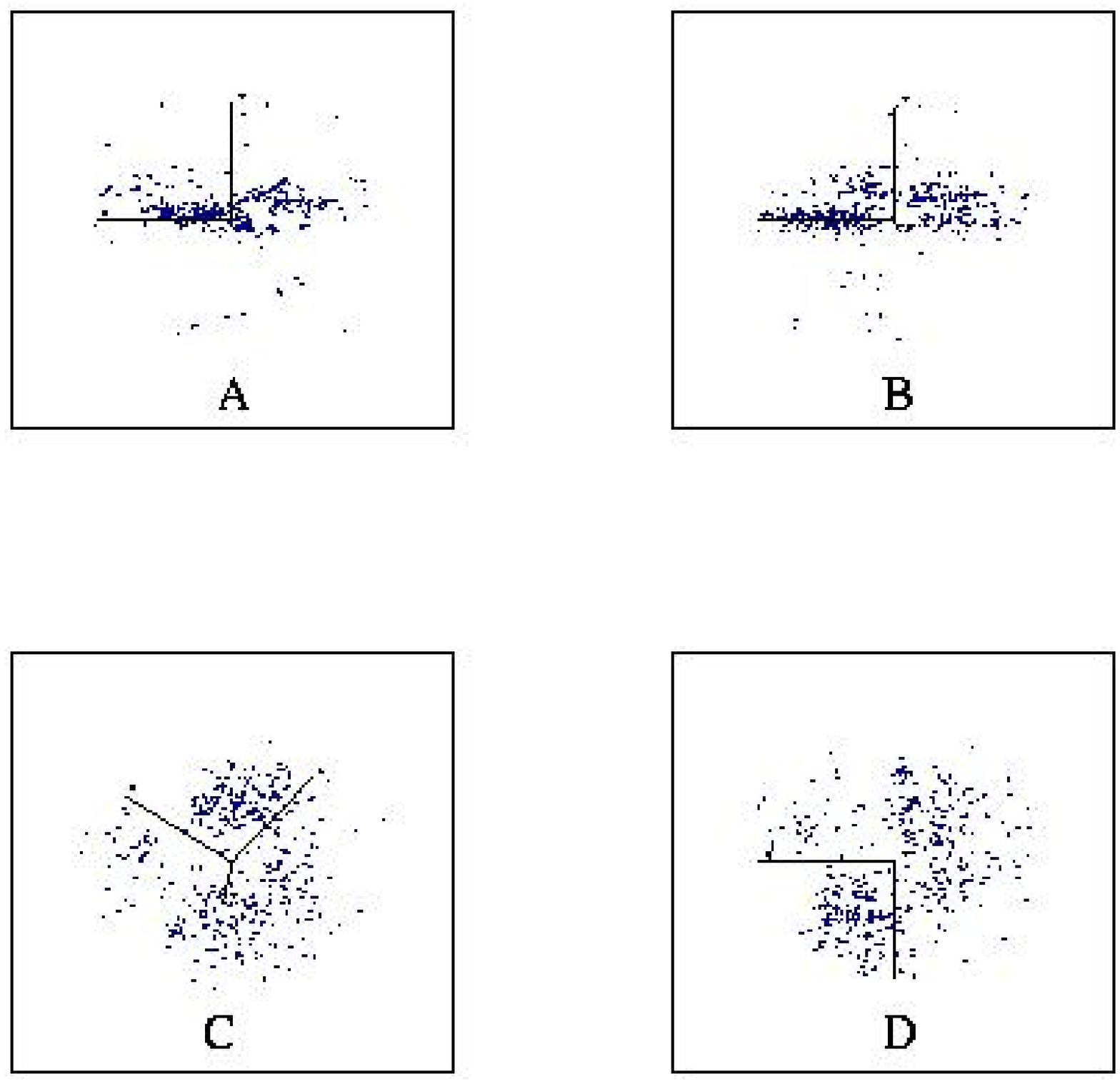}
\caption{Three-dimensional view of the 1665-MHz OH masers in Orion-KL, 
according to the kinematic model that was fitted (Section~\ref{model}).  The $x-$ and $y-$axes 
correspond to RA and Dec, while the $z-$axis 
is in the line-of-sight direction.  
Box A shows the view in the $x-y$ (RA--Dec) plane looking in the $+z-$direction, 
box B shows the view looking in the $+x-$direction (increasing RA), 
box C shows the view looking along the rotational axis of the best-fitting model, and box D shows the view looking down the $y-$axis (decreasing Dec.).}
\label{3dview}
\end{center}
\end{figure*}

For our kinematic modelling we used only the 1665-MHz masers.  
The celestial coordinates and radial velocities of the OH masers give incomplete information on the three-dimensional source geometry and motions.  
The OH maser positions are essentially the $x$- and $y$-components of each maser spot, but the $z$-coordinate is unknown.  
The radial velocity of a maser is related to the unknown velocity field and the ($x,y,z$) coordinates of the maser.  
We modelled the source using a simple kinematic model which was fitted to the data using the Control Random Search technique (Price 1976).  The model enables us to find the  $z$-coordinate for each maser from the radial velocity and the velocity field, and so produce a three-dimensional view of the OH maser distribution (Fig.~\ref{3dview}).  

We assumed, for modelling purposes, that IRc2 is at the kinematic centre.  However the results would not be affected if radio source I were taken as the centre.   We estimate that the kinematic centre of the OH masers is uncertain to $\sim$1~arcsec.  The masers were assumed to have a uniform expansion velocity $V_{\rm exp}$ away from IRc2, centred on an unknown radial velocity  $V_{\rm z0}$, plus solid body rotation corresponding to $V_{\rm rot}$ at 1-arcsec separation from IRc2.  The rotation axis is tilted by an angle $\alpha$ about the $y$-axis and an angle $\beta$ about the $x$-axis from the $x-z$ plane (following Reid et al. 1988).  The parameters $V_{\rm exp}$,  $V_{\rm z0}$,  $V_{\rm rot}$,  $\alpha$ and $\beta$ were the five unknowns.  For each maser and each position in the five-dimensional parameter space, we compared the observed and calculated radial velocity  $V_{\rm lsr}$ and found the $z$-position where the absolute value of the residual was minimized (allowing the maser to lie anywhere along the $z$-axis, with a distance corresponding to 30~arcsec of IRc2).    
The five-dimensional parameter space was searched, within a reasonable domain, using a random number generator to sample the parameters.  
The position in five-parameter space where the sum of the (absolute) velocity residuals is minimized was thereby located.  The best fitting model has the parameter values given in Table~\ref{parameters}.  
The angles correspond to a rotation axis inclined at 58\degr to the line-of-sight with a position angle on the sky of --35\degr.  A disc at this orientation is plotted in Fig.~\ref{ra-deccore}, where it fits neatly inside the  `hole' in the OH maser distribution that is centred on source I.  

\begin{figure}
\begin{center}
\vspace{75mm}
\includegraphics{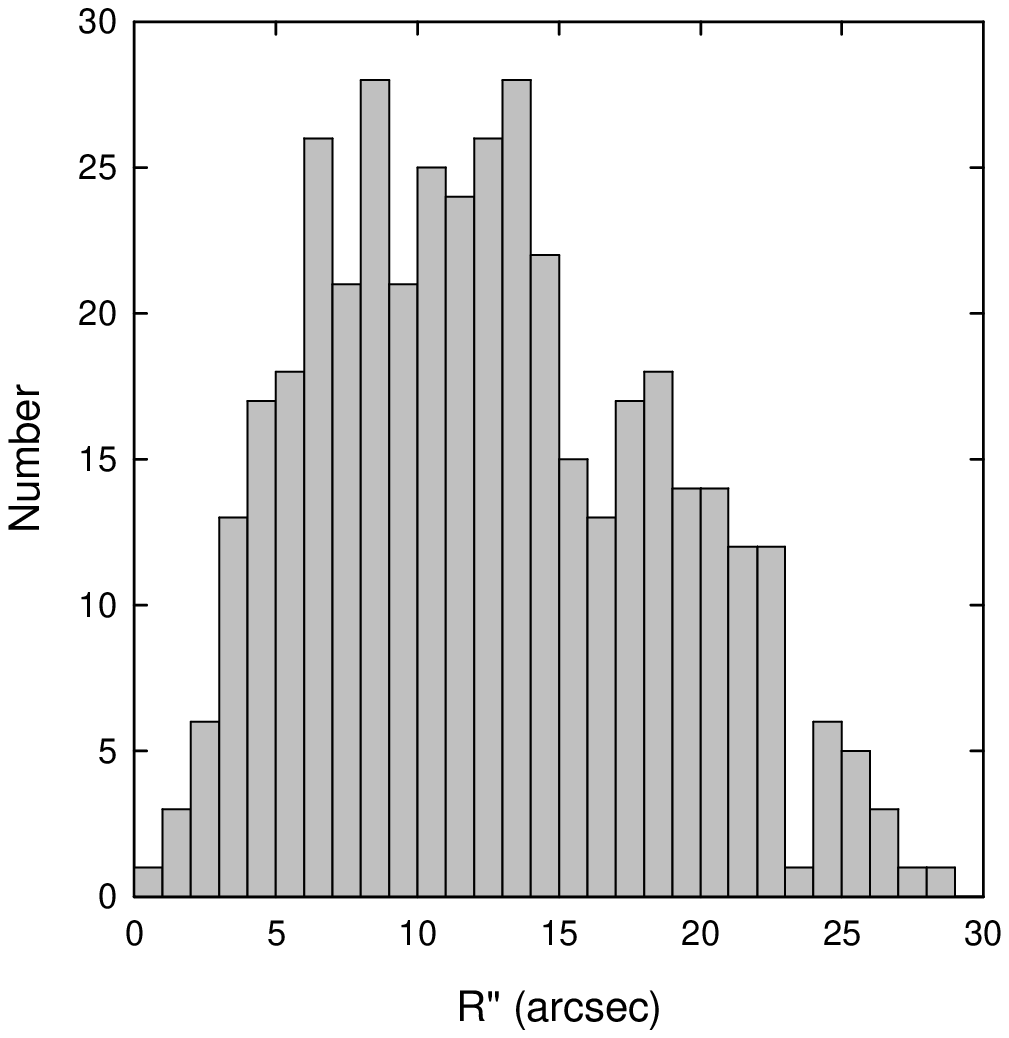}
\caption{Radial distribution of the 1665-MHz OH masers about the disc rotation axis, determined using the kinematic model  (Section~\ref{model}).}
\label{radial}
\end{center}
\end{figure}

Using the kinematic model we constructed views of the three-dimensional distribution of OH 1665-MHz masers, which are shown in Fig.~\ref{3dview}.  The masers lie in an irregularly filled torus, at radial distances ranging from 430~au to 13200~au, with a mean radius of 6200~au.  There is a well defined inner cavity, of radius $\sim$1300~au.  This cavity corresponds spatially with the 
SiO flared disc region mapped by Wright et al. (1995, 1996), which has a radial velocity range of $-10$ to +20~km~s$^{-1}$ that is roughly consistent with the range of $-12$ to  +30~km~s$^{-1}$ given by our kinematic model.  
The distribution in radial distance R\arcsec \hspace{0.5mm} from the rotation axis is plotted in Fig.~\ref{radial}.  
The distribution in the $z''$-direction, parallel to the rotation axis, has a full width to half maximum of 6000~au, and a total extent of 12000~au.

\begin{table}
\caption{Parameters of kinematic model.}
\label{parameters}

\begin{tabular}{lcc}
\hline
Parameter & Search Range & Best fit \\
\hline
 $V_{\rm exp}$  & 15--32~km~s$^{-1}$  & 21.0$\pm$3.5~km~s$^{-1}$  \\  
 $V_{\rm z0}$   & 8--10~km~s$^{-1}$   & 9.0$\pm$0.5~km~s$^{-1}$   \\
 $V_{\rm rot}$  & 1--5~km~s$^{-1}$    & 2.9$\pm$0.9~km~s$^{-1}$   \\
 $\alpha$       & 0--180\degr         & 46\degr$\pm$22\degr       \\
 $\beta$        & 0--90\degr          & 48\degr$\pm$19\degr       \\ 
\hline
\end{tabular}
\end{table}

The best-fitting expansion velocity is similar to that of the `low-velocity outflow' seen in H$_{2}$O masers (Genzel et al., 1981).  
Solid body rotation becomes the dominant motion beyond a distance of 3240~au from the centre (a distance corresponding to 7.2~arcsec).  Rotation is therefore the dominant motion for most of the system of OH masers, with however a significant component of expansion.  
The rotational period in our model is 760~yr, while the expansion timescale varies from $\sim$300~yr for the inner edge of the maser cavity to $\sim$3000~yr for the masers most distant from the expansion centre.

Most of the Orion nebula HH--objects are found
at the tips of the H$_{2}$ `fingers' in the bipolar outflow
shown in Fig.~\ref{subaru} and appear to have their origin in a newly created
massive star that gives rise to the infrared source IRc 2 in the
10$^{5}$ L$_{\odot}$ Orion-BN/KL nebula in the OMC--1 cloud core.  It has also been suggested that a minority of the prominent HH--objects have their origin in the neighbouring star forming region OMC--1S (Smith et al. 2004).  
However, excluding these few, Doi, O'Dell \& Hartigan (2002) have used HST proper-motion
measurements of the global expansion of the system of Orion HH--objects to 
demonstrate that the bipolar outflow is of Hubble type, i.e. the flow velocity is proportional to distance from the expansion centre. The bipolar flow  therefore consists of `bullets'
all ejected around a 1000 yr ago in a single explosive event.  Thus the expansion timescales determined for the OH masers bracket this H$_{2}$ timescale.  
These results suggest that the OH maser disc acquired its expansional motions in the same event that produced the  H$_{2}$ bullets and the HH--objects found at the tips of the  H$_{2}$ `fingers' in the bipolar outflow
shown in Fig. 7.

The orientation of the rotation axis determined here agrees with that of the molecular outflow traced in CO and H$_{2}$ (e.g. Chernin \& Wright 1996;  Kaifu et al. 2000).  It also coincides with the direction of the proper motions of source I and BN, and is close to the large-scale magnetic field direction inferred from far-infrared and submillimetre polarization (Schleuning 1998 and references therein).      
  
We note that the velocities of order 20~km~s$^{-1}$ implied by our kinematic model correspond to proper motions $\sim$9~mas~yr$^{-1}$, which could be measured by MERLIN within ten years.

\subsection{The Nature of Stream A and Arc B}
\label{arcs}

The major streams or arcs found in Orion-BN/KL have angular sizes greater than 5~arcsec, corresponding to projected sizes of at least 2250~au, which are almost unprecedented for interstellar OH masers.  
Stream A has no obvious counterpart in the infrared.  However it points in the direction of the dominant infrared source BN, in the direction of the radio proper motions of BN and source I.  Stream A has the appearance of a vapour trail that has formed in the wake of the runaway star BN.  From the proper motion of BN and the projected separation from the Northern tip of Stream A we can infer a timescale of 200~yr.  This may correspond to a cooling time, a formation time for OH, or a time for population inversion to be established following the passage of BN.   
   
Fig.~\ref{ra-dec} shows that the proper motions of BN and sources I and n project back to a position on the sky near the base of Stream A, in a region largely clear of OH masers.  This position is significantly displaced $\sim$4~arcsec Northwest of the centre of the OH maser torus.  
It has been suggested that BN and sources I and n originally belonged to a multiple stellar system that disintegrated $\sim$500~yr ago.  The disintegration could have due to a close dynamical interaction, as suggested by G\'{o}mez et al. (in press), or it could have been due to 
a merger event that also produced the molecular outflow, as suggested by Bally \& Zinnecker (2005).  Alternatively, Tan (2004) has proposed that BN was ejected from the  $\theta^{1}$ Ori C system $\sim$4000~yr ago.  Our data do not distinguish unambiguously between these possibilities.  However the fact that the large-scale OH maser torus is centred on source I slightly favours Tan's scenario, with BN being a passing runaway from  the  $\theta^{1}$ Ori C system.  

The other large-scale feature Arc B encloses IRc6, which is one of the cooler mid-IR sources, with polarization that is consistent with external illumination and with no evidence of embedded bright stars (Shuping et al. 2004 and references therein).  This suggests that Arc B might be associated with a photodissociation zone around IRc6.  In further support of this there is strong emission from CH$_{3}$CH at the position of IRc6, implying densities of at least 10$^{6}$~cm$^{-3}$ (Wilner, Wright \& Plambeck 1996).  

The large-scale maser structures found in Orion-BN/KL are similar in scale to OH maser structures recently reported in W3(OH).  Wright et al. (2004a,b) noted several arcs of 18-cm masers, which they interpreted as propagating shocks.  Harvey-Smith \& Cohen (2005) found a low brightness filament of excited OH 4765-MHz emission stretching for $\sim$2200~au and clearly related in its morphology to some of the ground-state OH arcs at 18-cm wavelength.  
The structures seen in Orion are similar in linear scale.  
The individual maser spots in Stream A and Arc B are well separated on the sky for the most part, but are nevertheless likely to be simply the high-gain cores of an extended stream of maser emission.  The typical flux densities of $\sim$0.1~Jy correspond to brightness temperatures of at least $\sim$3$\times$10$^{5}$~K.  

All these large-scale features in Orion and W3(OH) are relatively weak and therefore below the sensitivies of earlier surveys, for the most part.  Deeper observations of other well-known OH maser sources are needed to establish just how common such features are.

\subsection{The 1612-MHz Masers}
\label{1612}

The OH 1612-MHz masers have a more widespread distribution than the other OH masers, and different kinematics (Section~\ref{kinematics}).  This suggests that they trace a region with different physical conditions.  
Gray, Field \& Doel (1992) give the following conditions for strong 1612-MHz masers:  
molecular hydrogen density $n_{\rm H_{2}}$=\(6\times 10^{6}\)~cm$^{-3}$,  
$n_{\rm OH}$ 60~cm$^{-3}$,  gas kinetic temperature $T_{\rm k}$=30~K,  dust temperature at maser site $T_{\rm d}$  30~K, external radiation field  
$T_{\rm x}$  80~K  and velocity shift 
$\Delta V$ 2.0~km\,s$^{-1}$.  
This is cooler than typical for OH masers, but is consistent with the location of these particular 1612-MHz masers further than normal from the main source of infrared luminosity.  The velocity gradient needed for strong maser action is also consistent with the association of these masers with shocked gas in the molecular outflow (Section~\ref{nearIR}).   

\begin{figure}
\begin{center}
\vspace{83mm}
\includegraphics{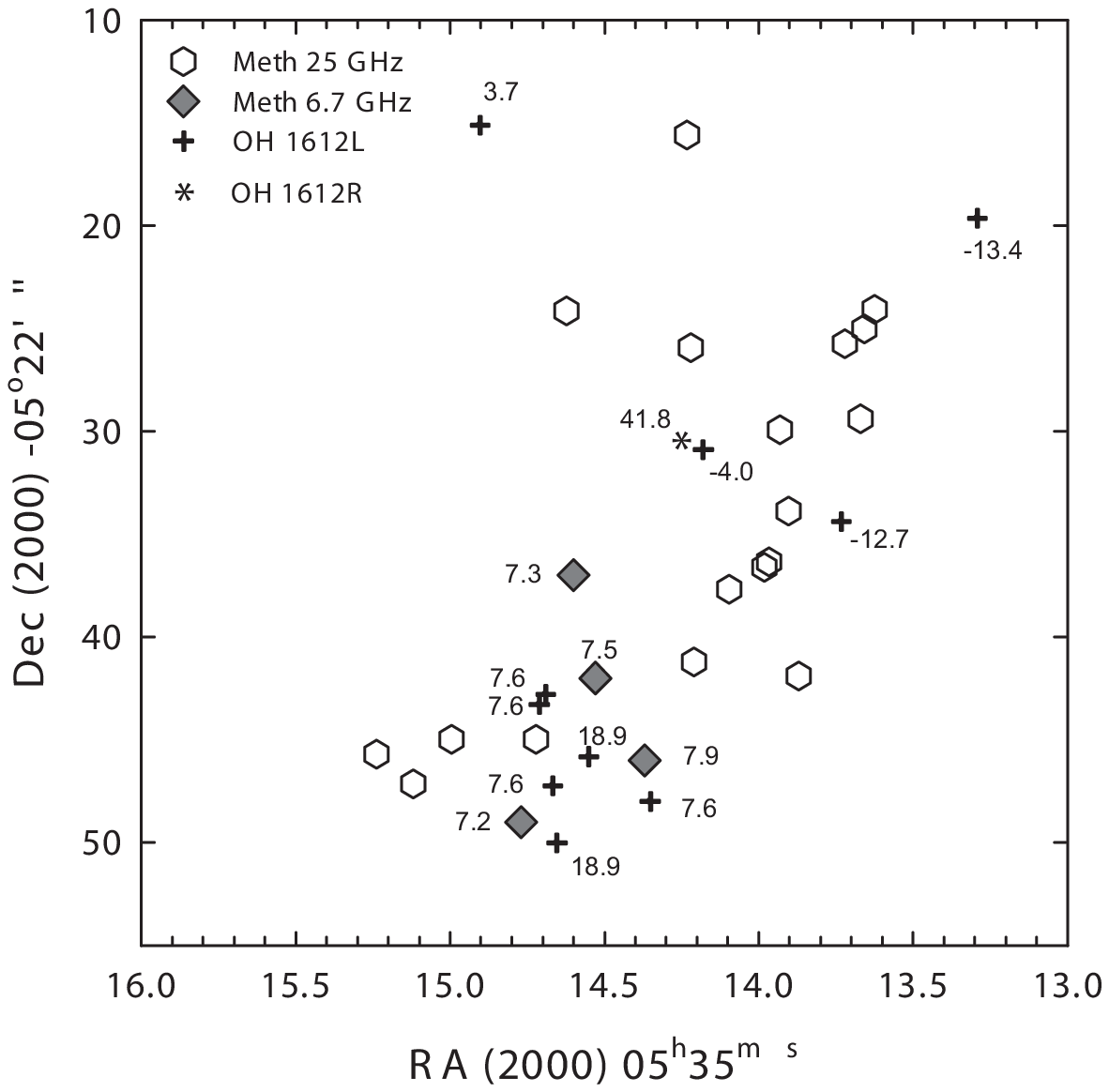}
\caption{Positions and velocities of OH 1612-MHz masers (this paper) compared with those of the newly discovered methanol 6.7-GHz masers (Voronkov et al. 2005).  Positions of 25-GHz methanol masers from Johnston et al. (1997) and Johnston et al. (1992) are also shown.  The radial velocities at 25~GHz are confined to a narrow range of 7--10~km~s$^{-1}$. }
\label{methanol}
\end{center}
\end{figure}

We note a further possible positional association between the 1612-MHz masers and the methanol 6.7-GHz masers recently discovered by Voronkov et al. (2005).  In Fig.~\ref{methanol} we show the positions and velocities of the two species.  There is a clear region of overlap centred around 
 RA (J2000)~=~05$^{\rm h}$35$^{\rm m}$14\fs5, Dec. (J2000)~=~-05\degr22\arcmin45\farcs9 and $V_{\rm lsr}$=~+7.5~km~s$^{-1}$.  The 6.7-GHz positions have errors of $\sim$2~arcsec.  It will be important to make 6.7-GHz measurements of higher precision to examine this correspondence more closely.  The positions of 25-GHz masers from Johnston et al. (1997, 1992) are also shown in the figure for completeness.  Only one of these falls in the region of interest.  We note that there is no association with the OH mainline masers.  This is consistent with the 25-GHz masers being class I masers, which are thought to be collisionally pumped (cf. Cragg et al. 2002).   

The 6.7-GHz maser transition is the prototype class II maser, usually thought to be radiatively pumped.  
The 6.7-GHz masers in Orion are of therefore of interest because of their apparent association with 25-GHz masers.  
Voronkov et al. have considered the pumping requirements for the coexistence of both types of methanol maser and find that both types can occur simultaneously at low temperature ($\sim$60~K) and low molecular hydrogen density ($\sim$10$^{5}$~cm$^{-3}$).  These conditions are not too dissimilar to those needed for strong 1612-MHz masers, as given earlier.  In summary, it appears that in Orion we may have the first examples of both methanol class II and OH ground state masers located far from the main source of infrared luminosity, and associated instead with the interaction between the molecular outflow and the surrounding gas.

\subsection{Proplyd region}
\label{proplyds2}

The proplyds originally designated LV~1--6 within the Orion
Trapezium cluster are those most likely to exhibit maser emission
for they are irradiated by the intense  field of the
O6.5 star  $\theta^{1}$ Ori~C. Of these LV~1, now
resolved as a binary proplyd (168--326 NW \& SE) by
O'Dell \& Wen (1994), and LV~2 (proplyd 167-317 of   O'Dell \& Wen 1994)
have been most intensively observed (Graham et al. 2002; Henney et al. 2002
respectively).

The 8$\times$10$^{14}$ cm radius LV~2 was shown by
Henney et al. (2002) to be irradiated by a flux of 3$\times$10$^{4}$ 
cm$^{-2}$ s$^{-1}$ Lyman photons. These produce an ionised skin with
a measured electron density n$_{e}$~=~2.6$\times$10$^{6}$ cm$^{-3}$
on the surface of a neutral disk or cocoon which envelopes the low
mass YSO (Meaburn 1988).  From 1.3-mm interferometric
observations Bally et al. (1998b) deduce an upper limit to the mass
of molecular hydrogen in this neutral component of $\sim$0.015~M$_{\odot}$ 
to give an H$_{2}$ density of n$_{o}$~$\leq$~4.2$\times$10$^{9}$ cm$^{-3}$.
On the other hand, simple pressure balance between the ionised and neutral gas
with temperatures of T$_{e}$ = 10$^{4}$ K and T$_{o}$ = 10$^{3}$ K would
give n$_{o}\approx$~5.2$\times$10$^{9}$ cm$^{-3}$.  

OH masers have yet to be found in regions of purely low-mass star-formation.  
However, it is this dense, circumstellar molecular gas around proplyds, 
stimulated by photons from   $\theta^{1}$Ori~C, that is the potential  source of any
maser emission.  We would expect to find OH masers at a distance of $\sim$7$\times$10$^{17}$cm from an O6.5 star (cf. Baart \& Cohen 1985).  
However, if Keplerian motion around the 
YSO occurs in this neutral cocoon it will have a rotational velocity
of 4~km~s$^{-1}$
at its outer edge with higher velocities towards the centre
(proportional to radius$^{-1/2}$). In these 
circumstances the velocity coherence required for strong maser amplification
will occur over a path length that is much smaller than the overall radius of the proplyd.  
This, together with possible chemical effects (Section~\ref{proplyds1}), could explain our non-detection of OH masers in this region.  

\section{Conclusions}

The distribution of OH masers in the Orion-BN/KL region is far more extensive than previously realized, covering a region of 30~arcsec extent and a radial velocity range from --13 to +42~km~s$^{-1}$.  The bulk of the emission can be modelled in terms of a rotating and expanding torus, centred on IRc2 or radio source I, with an inner cavity of $\sim$1300~au radius.  The rotation axis has the same position angle and inclination to the line-of-sight as the molecular outflow and the large scale magnetic field inferred from mm- and submm-polarization (Section~\ref{model}).  The dynamical timescale is similar to that of the explosive event that produced the widespread shocked H$_{2}$ emission.  It is likely that the OH masers trace the interaction between the low-velocity molecular outflow and the molecular hot core, and that they acquired their expansional motions in the same event that produced the outflow.  

Of particular interest is a string of masers, Stream A, at $\sim$21~km~s$^{-1}$, that extends at position angle $\sim$45\degr between IRc2 and BN, in the direction of the radio proper motions of these two dominant sources.  We suggest that Stream A may have appeared, like a vapour trail, in the wake of the runaway star BN.  The proper motions of BN and sources I and n project back to the base of Stream A (Fig.~\ref{ra-dec}), $\sim$4~arcsec Northwest of the centre of the OH maser torus, a position that is largely devoid of masers.

The 1612-MHz masers have a more widespread distribution than the other OH masers, with kinematics that are more like those of the H$_{2}$O masers associated with the outflow.  Many of these 1612-MHz masers are spatially associated with fingers of shocked H$_{2}$ emission (Fig.~\ref{subcore}).  These OH masers are thought to require relatively low gas and dust temperatures for their inversion (Section~\ref{1612}).  Apart from the OH 1612-MHz masers, the other OH masers have complementary distributions and kinematics to the  H$_{2}$O masers (Figs.~\ref{ra-dec}--\ref{dec-vel}), with essentially no overlap.  OH is also spatially and kinematically distinct from the class I methanol 25-GHz masers.  A possible correspondence between OH 1612-MHz and class II methanol 6.7-GHz masers reported by Voronkov et al. (2005) requires further study at higher angular resolution. 

The magnetic field strength in the OH maser regions ranges from 1.8 to 16.3~mG, with a possible field reversal (Section~\ref{zeeman}).

A search for OH masers associated with the proplyds around  $\theta^{1}$ Ori C yielded only upper limits, which may indicate a low OH abundance in this more evolved region.

\vspace{1.2cm}
\noindent \large {\bf ACKNOWLEDGMENTS}\normalsize
\vspace{0.5cm}

We thank the anonymous referee for helpful comments and in particular for 
drawing our attention to the paper by G\'{o}mez et al.
We thank Busaba Hutawarakorn for assistance with the data reduction,  
Ralph Shuping for providing a high resolution version of the Keck 
image in Fig.~5, and Masa Hayashi for providing high resolution 
Subaru data used in Figs. 6 and 7.      
NG gratefully acknowledges the Thai Astronomy Co-operative Research
Network  for their support.  
MERLIN is a national facility operated by the University of Manchester
on behalf of PPARC.

\newpage

\setcounter{table}{0}
\begin{table*}
\caption{OH 1.6-GHz Masers in Orion-BN/KL.}
\label{OHtab}

\begin{tabular}{cccccccc}
\hline
Transition &  Feature &  Velocity   &  RA(1950)   &  Dec(1950) &  RA(2000) & Dec(2000) & Intensity  \\ 
           &  No.     &  (km~s$^{-1}$) &  05$^{\rm h}$32$^{\rm m}$  $^{\rm s}$ & --5$^{\circ}$:24\arcmin:  \arcsec &  05$^{\rm h}$h35$^{\rm m}$ $^{\rm s}$  &  --5$^{\circ}$:22\arcmin:  \arcsec & (Jy~beam$^{-1}$) \\
\hline 

{\em 1612L} & 1  & --13.41 & 45.8137 & 13.153 & 13.2911 & 19.624 & 0.20  \\
            & 2 & --12.68 & 46.2592 & 27.890 & 13.7318 & 34.393 & 0.11  \\
         & 3 & --3.97 & 46.7062 & 24.344 &  14.1800 & 30.880 & 0.20  \\
        &  4 &  3.66 & 47.4242 & 08.524 &  14.9030 & 15.112 & 0.11  \\
        &  5 &  7.64 & 46.8839 & 41.471 &  14.3521 & 48.019 & 0.09  \\
        & & & & & & \\
         & 6 &  7.64 & 47.1993 & 40.678 &  14.6677 & 47.249  & 0.12  \\
         & 7 &  7.64 & 47.2210 & 36.224 &  14.6908 & 42.797  & 0.12  \\
        &  8 &  7.64 & 47.2417 & 36.726 &  14.7114 & 43.300  & 0.17  \\ 
        &  9 & 18.93 & 47.0810 & 39.263 &  14.5499 & 45.826  & 0.08 \\ 
        & 10 & 18.93 & 47.1858 & 43.452 &  14.6533 & 50.022  & 0.13  \\ 
\hline
{\em 1612R}      
 &  1 & 41.80 & 46.7773 & 24.354 &  14.2510 & 30.895  & 0.12  \\ 
\hline
{\em 1665L}  
 & 1 & --10.70 & 46.6328 & 22.505 & 14.1072 & 29.035 & 0.14 \\
 & 2 & --10.42 & 46.7912 & 22.176 & 14.2656 & 28.718 & 0.24 \\
 & 3 &  --9.64 & 46.8095 & 22.013 & 14.2840 & 28.557 & 0.14 \\
 & 4 &  --9.64 & 47.0398 & 22.993 & 14.5139 & 29.553 & 0.34 \\
 & 5 &  --9.32 & 46.4472 & 22.867 & 13.9214 & 29.384 & 0.17 \\
         & & & & & & \\
 & 6 &  --9.11 & 46.7827 & 21.337 & 14.2574 & 27.878 & 0.13 \\
 & 7 &  --8.59 & 46.8077 & 23.509 & 14.2817 & 30.052 & 0.27 \\
 & 8 &  --8.24 & 46.4518 & 22.571 & 13.9261 & 29.088 & 0.12 \\
 & 9 &  --8.02 & 46.7338 & 21.544 & 14.2084 & 28.082 & 0.17 \\
 & 10 & --7.88 & 46.6340 & 23.458 & 14.1080 & 29.989 & 0.24 \\
         & & & & & & \\
 & 11 & --7.88 & 46.7335 & 21.526 & 14.2082 & 28.064 & 0.17 \\
 & 12 & --7.88 & 46.6316 & 22.036 & 14.1061 & 28.566 & 0.20 \\
 & 13 & --7.53 & 46.9440 & 20.729 & 14.4189 & 27.282 & 0.10 \\
 & 14 & --7.53 & 46.9255 & 20.468 & 14.4005 & 27.019 & 0.11 \\
 & 15 & --7.36 & 46.8692 & 20.569 & 14.3442 & 27.116 & 0.90 \\
         & & & & & & \\
 & 16 & --7.36 & 46.9485 & 20.810 & 14.4234 & 27.364 & 0.13 \\
 & 17 & --6.89 & 46.7149 & 21.956 & 14.1894 & 28.493 & 0.16 \\
 & 18 & --6.83 & 46.7282 & 20.943 & 14.2031 & 27.480 & 0.12 \\
 & 19 & --6.61 & 46.7719 & 21.301 & 14.2466 & 27.842 & 0.19 \\
 & 20 & --6.13 & 46.7809 & 20.927 & 14.2558 & 27.468 & 0.17 \\
         & & & & & & \\
 & 21 & --6.13 & 46.7442 & 20.854 & 14.2191 & 27.392 & 0.10 \\
 & 22 & --5.83 & 46.8968 & 23.735 & 14.3707 & 30.284 & 0.73 \\
 & 23 & --5.77 & 46.8065 & 21.846 & 14.2811 & 28.389 & 0.10 \\
 & 24 & --5.77 & 46.7768 & 21.284 & 14.2516 & 27.825 & 0.14 \\
 & 25 & --5.77 & 46.9090 & 23.745 & 14.3829 & 30.295 & 0.42 \\
         & & & & & & \\
 & 26 & --5.69 & 46.7619 & 20.790 & 14.2368 & 27.329 & 0.26 \\
 & 27 & --5.42 & 46.9295 & 20.677 & 14.4044 & 27.229 & 0.10 \\
 & 28 & --5.42 & 46.9134 & 23.784 & 14.3873 & 30.335 & 0.69 \\
 & 29 & --5.42 & 46.9386 & 33.358 & 14.4094 & 39.910 & 0.08 \\
 & 30 & --4.37 & 46.8096 & 21.692 & 14.2842 & 28.236 & 0.09 \\
         & & & & & & \\
 & 31 & --4.02 & 46.7662 & 21.057 & 14.2410 & 27.597 & 0.14 \\
 & 32 & --3.89 & 47.1898 & 17.315 & 14.6658 & 23.886 & 0.28 \\
 & 33 & --3.31 & 46.5602 & 21.601 & 14.0348 & 28.126 & 0.16 \\
 & 34 & --2.96 & 46.9105 & 19.907 & 14.3857 & 26.458 & 0.15 \\
 & 35 & --2.96 & 46.5629 & 21.625 & 14.0375 & 28.150 & 0.13 \\
         & & & & & & \\
 & 36 & --2.61 & 47.0183 & 33.522 & 14.4890 & 40.080 & 0.10 \\
 & 37 & --2.26 & 46.7894 & 20.575 & 14.2644 & 27.117 & 0.21 \\
 & 38 & --1.91 & 46.8197 & 20.397 & 14.2948 & 26.941 & 0.22 \\
 & 39 & --1.56 & 46.6744 & 21.334 & 14.1491 & 27.867 & 0.35 \\
 & 40 & --1.56 & 46.8823 & 20.338 & 14.3574 & 26.887 & 0.12 \\
         & & & & & & \\
 & 41 & --1.38 & 46.9594 & 23.707 & 14.4334 & 30.261 & 0.44 \\
\hline
\end{tabular}
\end{table*}

\begin{table*} 
\addtocounter{table}{-1}
\caption{OH 1.6-GHz Masers in Orion-BN/KL, continued.}
\begin{tabular}{cccccccc}
\hline
Transition &  Feature &  Velocity   &  RA(1950)   &  Dec(1950) &  RA(2000) & Dec(2000) & Intensity  \\ 
           &  No.     &  (km~s$^{-1}$) &  05$^{\rm h}$32$^{\rm m}$  $^{\rm s}$ & --5$^{\circ}$:24\arcmin:  \arcsec &  05$^{\rm h}$h35$^{\rm m}$ $^{\rm s}$  &  --5$^{\circ}$:22\arcmin:  \arcsec & (Jy~beam$^{-1}$) \\
\hline

{\em 1665L} 
 & 42 & --0.85 & 46.8218 & 20.452 & 14.2968 & 26.997 & 0.42 \\
 & 43 &  0.30 & 46.8021 & 23.549 & 14.2761 & 30.092 & 0.58 \\
 & 44 &  0.37 & 47.1572 & 19.302 & 14.6326 & 25.870 & 0.29 \\
 & 45 &  0.46 & 46.9636 & 23.523 & 14.4376 & 30.077 & 1.84 \\
         & & & & & & \\
 & 46 &  1.14 & 46.9246 & 22.479 & 14.3989 & 29.031 & 0.24 \\
 & 47 &  1.61 & 46.9754 & 23.672 & 14.4493 & 30.228 & 0.13 \\
 & 48 &  1.61 & 46.9424 & 23.138 & 14.4165 & 29.691 & 0.13 \\
 & 49 &  1.61 & 46.9523 & 25.187 & 14.4258 & 31.740 & 0.13 \\
 & 50 &  1.84 & 46.9463 & 23.766 & 14.4202 & 30.319 & 6.77 \\
         & & & & & & \\
 & 51 &  2.99 & 46.9400 & 22.603 & 14.4142 & 29.156 & 1.25 \\
 & 52 &  3.37 & 46.9964 & 23.444 & 14.4704 & 30.001 & 0.12 \\
 & 53 &  3.37 & 46.5059 & 20.795 & 13.9808 & 27.317 & 0.11 \\
 & 54 &  3.65 & 46.8365 & 20.701 & 14.3114 & 27.246 & 0.26 \\
 & 55 &  3.70 & 46.9524 & 23.681 & 14.4263 & 30.235 & 1.25 \\
         & & & & & & \\
 & 56 &  3.72 & 46.4051 & 20.615 & 13.8801 & 27.129 & 0.11 \\
 & 57 &  3.72 & 46.3534 & 21.314 & 13.8281 & 27.825 & 0.10 \\
 & 58 &  3.72 & 46.7119 & 20.664 & 14.1868 & 27.201 & 0.19 \\
 & 59 &  3.89 & 46.5796 & 21.329 & 14.0543 & 27.855 & 0.57 \\
 & 60 &  4.03 & 46.6630 & 20.911 & 14.1379 & 27.444 & 0.51 \\
         & & & & & & \\
 & 61 &  4.07 & 46.5423 & 21.241 & 14.0171 & 27.765 & 0.14 \\
 & 62 &  4.63 & 46.7483 & 20.876 & 14.2232 & 27.415 & 1.22 \\
 & 63 &  4.69 & 46.6243 & 21.026 & 14.0991 & 27.555 & 0.35 \\
 & 64 &  4.77 & 46.9233 & 23.267 & 14.3974 & 29.819 & 0.16 \\
 & 65 &  4.77 & 46.6881 & 20.622 & 14.1631 & 27.157 & 0.15 \\
         & & & & & & \\
 & 66 &  4.77 & 46.9581 & 22.943 & 14.4323 & 29.497 & 0.12 \\
 & 67 &  4.77 & 46.5770 & 21.333 & 14.0518 & 27.859 & 0.10 \\
 & 68 &  4.77 & 46.4888 & 21.276 & 13.9636 & 27.796 & 0.10 \\
 & 69 &  4.77 & 46.7341 & 20.653 & 14.2091 & 27.191 & 0.14 \\
 & 70 &  4.94 & 46.9280 & 24.075 & 14.4018 & 30.627 & 13.19 \\
         & & & & & & \\
 & 71 &  5.58 & 46.9628 & 22.884 & 14.4370 & 29.438 & 1.35 \\
 & 72 &  5.83 & 46.6159 & 20.657 & 14.0909 & 27.186 & 0.11 \\
 & 73 &  5.83 & 46.3649 & 18.674 & 13.8405 & 25.185 & 0.10 \\
 & 74 &  5.95 & 46.6308 & 21.075 & 14.1056 & 27.605 & 0.17 \\
 & 75 &  6.53 & 46.9632 & 22.822 & 14.4374 & 29.376 & 0.13 \\
         & & & & & & \\
 & 76 &  6.64 & 46.4056 & 18.590 & 13.8812 & 25.104 & 0.56 \\
 & 77 &  6.66 & 46.9574 & 23.654 & 14.4313 & 30.207 & 3.54 \\
 & 78 &  6.88 & 47.4785 & 19.472 & 14.9538 & 26.063 & 0.11 \\
 & 79 &  7.29 & 46.9105 & 24.319 & 14.3842 & 30.870 & 2.03 \\
 & 80 &  7.42 & 46.9717 & 20.929 & 14.4466 & 27.484 & 0.23 \\
         & & & & & & \\
 & 81 &  7.59 & 46.5247 & 20.806 & 13.9996 & 27.329 & 0.12 \\
 & 82 &  7.71 & 46.6617 & 29.943 & 14.1337 & 36.475 & 0.20 \\
 & 83 &  7.70 & 46.8017 & 24.012 & 14.2755 & 30.555 & 0.26 \\
 & 84 &  7.93 & 47.4563 & 19.610 & 14.9316 & 26.200 & 0.55 \\
 & 85 &  8.36 & 46.8554 & 25.968 & 14.3286 & 32.514 & 0.31 \\
         & & & & & & \\
 & 86 &  8.91 & 46.7967 & 24.106 & 14.2705 & 30.648 & 2.99 \\
 & 87 &  8.98 & 46.9087 & 24.312 & 14.3824 & 30.863 & 1.57 \\
 & 88 &  8.99 & 46.9060 & 22.794 & 14.3802 & 29.344 & 0.11 \\
 & 89 &  9.06 & 46.9558 & 20.654 & 14.4307 & 27.207 & 0.40 \\
 & 90 &  9.70 & 47.5408 & 20.189 & 15.0158 & 26.786 & 0.33 \\
         & & & & & & \\
 & 91 &  9.81 & 46.5539 & 20.811 & 14.0288 & 27.336 & 0.37 \\
 & 92 & 10.75 & 47.0620 & 24.863 & 14.5356 & 31.424 & 0.21 \\
 & 93 & 10.75 & 47.3672 & 21.696 & 14.8418 & 28.279 & 0.14 \\
\hline
\end{tabular}
\end{table*}

\begin{table*} 
\addtocounter{table}{-1}
\caption{OH 1.6-GHz Masers in Orion-BN/KL, continued.}
\begin{tabular}{cccccccc}
\hline
Transition &  Feature &  Velocity   &  RA(1950)   &  Dec(1950) &  RA(2000) & Dec(2000) & Intensity  \\ 
           &  No.     &  (km~s$^{-1}$) &  05$^{\rm h}$32$^{\rm m}$  $^{\rm s}$ & --5$^{\circ}$:24\arcmin:  \arcsec &  05$^{\rm h}$h35$^{\rm m}$ $^{\rm s}$  &  --5$^{\circ}$:22\arcmin:  \arcsec & (Jy~beam$^{-1}$) \\
\hline

{\em 1665L} 
 & 94 & 11.10 & 46.7601 & 30.912 & 14.2317 & 37.452 & 0.13 \\
 & 95 & 11.19 & 47.4566 & 18.226 & 14.9322 & 24.816 & 0.29 \\
         & & & & & & \\
 & 96 & 11.21 & 47.3960 & 18.029 & 14.8717 & 24.615 & 0.38 \\
 & 97 & 11.59 & 46.9193 & 23.131 & 14.3934 & 29.682 & 0.26 \\
 & 98 & 11.89 & 47.4229 & 21.920 & 14.8973 & 28.508 & 2.17 \\
 & 99 & 12.16 & 47.0610 & 24.863 & 14.5346 & 31.424 & 0.25 \\
 & 100 & 12.55 & 46.9741 & 23.885 & 14.4479 & 30.440 & 2.50 \\
         & & & & & & \\
 & 101 & 12.86 & 47.0105 & 21.980 & 14.4849 & 28.538 & 0.10 \\
 & 102 & 12.94 & 46.9106 & 23.520 & 14.3846 & 30.071 & 0.50 \\
 & 103 & 13.21 & 47.0528 & 21.892 & 14.5273 & 28.453 & 0.11 \\
 & 104 & 13.21 & 46.9949 & 21.645 & 14.4695 & 28.202 & 0.19 \\
 & 105 & 13.54 & 46.9486 & 23.990 & 14.4225 & 30.543 & 4.66 \\
         & & & & & & \\
 & 106 & 13.56 & 46.9987 & 21.742 & 14.4733 & 28.299 & 0.37 \\
 & 107 & 13.92 & 46.7596 & 19.790 & 14.2348 & 26.330 & 0.15 \\
 & 108 & 15.15 & 46.7672 & 19.970 & 14.2424 & 26.510 & 0.10 \\
 & 109 & 15.32 & 46.7762 & 30.758 & 14.2479 & 37.298 & 0.11 \\
 & 110 & 15.67 & 46.7458 & 19.896 & 14.2210 & 26.434 & 0.14 \\
         & & & & & & \\
 & 111 & 15.73 & 47.1206 & 22.445 & 14.5949 & 29.010 & 0.56 \\
 & 112 & 15.81 & 47.0033 & 21.966 & 14.4778 & 28.523 & 0.33 \\
 & 113 & 15.90 & 47.3826 & 21.547 & 14.8572 & 28.131 & 1.62 \\
 & 114 & 16.23 & 47.0147 & 22.161 & 14.4891 & 28.719 & 0.69 \\
 & 115 & 16.50 & 47.3545 & 21.873 & 14.8290 & 28.455 & 0.23 \\
         & & & & & & \\
 & 116 & 16.86 & 47.3909 & 20.129 & 14.8660 & 26.714 & 0.15 \\
 & 117 & 17.03 & 47.1812 & 21.727 & 14.6557 & 28.297 & 0.20 \\
 & 118 & 17.08 & 47.3752 & 21.588 & 14.8498 & 28.172 & 0.15 \\
 & 119 & 17.08 & 47.0071 & 21.910 & 14.4816 & 28.468 & 0.13 \\
 & 120 & 17.22 & 47.1108 & 21.999 & 14.5853 & 28.564 & 0.18 \\
         & & & & & & \\
 & 121 & 17.43 & 47.1021 & 22.817 & 14.5764 & 29.381 & 0.13 \\
 & 122 & 17.43 & 47.3406 & 22.604 & 14.8149 & 29.185 & 0.12 \\
 & 123 & 17.43 & 47.0877 & 22.617 & 14.5620 & 29.180 & 0.12 \\
 & 124 & 17.61 & 47.2131 & 22.777 & 14.6873 & 29.349 & 0.28 \\
 & 125 & 17.78 & 47.0585 & 22.618 & 14.5328 & 29.179 & 0.13 \\
         & & & & & & \\
 & 126 & 17.78 & 47.1081 & 22.726 & 14.5824 & 29.291 & 0.14 \\
 & 127 & 17.78 & 47.3509 & 22.000 & 14.8254 & 28.582 & 0.13 \\
 & 128 & 17.78 & 47.1612 & 22.569 & 14.6355 & 29.137 & 0.10 \\
 & 129 & 17.78 & 47.1809 & 21.706 & 14.6555 & 28.276 & 0.30 \\
 & 130 & 17.78 & 47.2869 & 22.111 & 14.7614 & 28.688 & 0.11 \\
         & & & & & & \\
 & 131 & 17.78 & 47.4061 & 21.989 & 14.8806 & 28.575 & 0.10 \\
 & 132 & 17.91 & 47.0200 & 21.993 & 14.4945 & 28.552 & 0.42 \\
 & 133 & 18.13 & 47.3239 & 21.992 & 14.7984 & 28.572 & 0.10 \\
 & 134 & 18.13 & 47.1849 & 23.365 & 14.6589 & 29.936 & 0.24 \\
 & 135 & 18.13 & 46.8631 & 20.019 & 14.3382 & 26.566 & 0.11 \\
         & & & & & & \\
 & 136 & 18.13 & 47.0591 & 22.609 & 14.5334 & 29.170 & 0.11 \\
 & 137 & 18.13 & 47.1708 & 23.045 & 14.6449 & 29.614 & 0.28 \\
 & 138 & 18.13 & 47.2031 & 22.800 & 14.6773 & 29.372 & 0.15 \\
 & 139 & 18.37 & 47.1359 & 22.256 & 14.6103 & 28.823 & 1.91 \\
 & 140 & 18.45 & 47.3202 & 22.860 & 14.7944 & 29.441 & 0.74 \\
         & & & & & & \\
 & 141 & 18.49 & 47.3692 & 21.736 & 14.8438 & 28.319 & 0.10 \\
 & 142 & 18.49 & 47.2276 & 22.656 & 14.7019 & 29.229 & 0.14 \\
 & 143 & 18.68 & 47.2708 & 23.175 & 14.7449 & 29.752 & 0.17 \\
 & 144 & 18.77 & 47.1887 & 23.174 & 14.6628 & 29.745 & 0.38 \\
 & 145 & 18.84 & 47.2506 & 23.248 & 14.7247 & 29.823 & 0.10 \\
\hline
\end{tabular}
\end{table*}

\begin{table*} 
\addtocounter{table}{-1}
\caption{OH 1665-MHz Masers in Orion-BN/KL, continued.}
\label{65tab}
\begin{tabular}{cccccccc}
\hline
Transition &  Feature &  Velocity   &  RA(1950)   &  Dec(1950) &  RA(2000) & Dec(2000) & Intensity  \\ 
           &  No.     &  (km~s$^{-1}$) &  05$^{\rm h}$32$^{\rm m}$  $^{\rm s}$ & --5$^{\circ}$:24\arcmin:  \arcsec &  05$^{\rm h}$h35$^{\rm m}$ $^{\rm s}$  &  --5$^{\circ}$:22\arcmin:  \arcsec & (Jy~beam$^{-1}$) \\
\hline 

{\em 1665L} 
 & 146 & 18.84 & 47.0446 & 22.294 & 14.5190 & 28.854 & 0.10 \\
 & 147 & 19.19 & 47.2460 & 23.949 & 14.7199 & 30.524 & 0.10 \\
 & 148 & 19.19 & 46.9805 & 21.779 & 14.4551 & 28.335 & 0.11 \\
 & 149 & 19.28 & 47.3395 & 22.238 & 14.8139 & 28.820 & 1.13 \\
 & 150 & 19.30 & 47.1008 & 22.783 & 14.5750 & 29.347 & 0.41 \\
         & & & & & & \\
 & 151 & 19.39 & 47.1682 & 22.963 & 14.6424 & 29.532 & 0.73 \\
 & 152 & 19.49 & 47.2982 & 22.553 & 14.7724 & 29.132 & 1.38 \\
 & 153 & 19.54 & 47.1921 & 18.773 & 14.6676 & 25.344 & 0.82 \\
 & 154 & 19.54 & 47.2170 & 23.168 & 14.6911 & 29.740 & 0.12 \\
 & 155 & 19.72 & 47.1863 & 22.905 & 14.6604 & 29.476 & 0.65 \\
         & & & & & & \\
 & 156 & 19.89 & 47.3449 & 22.840 & 14.8191 & 29.422 & 0.11 \\
 & 157 & 20.00 & 47.2990 & 22.998 & 14.7731 & 29.576 & 0.39 \\
 & 158 & 20.06 & 47.2813 & 22.322 & 14.7557 & 28.900 & 1.34 \\
 & 159 & 20.22 & 47.2210 & 23.460 & 14.6950 & 30.033 & 0.38 \\
 & 160 & 20.24 & 47.1731 & 22.703 & 14.6474 & 29.273 & 0.24 \\
         & & & & & & \\
 & 161 & 20.24 & 47.3741 & 21.408 & 14.8487 & 27.992 & 0.20 \\
 & 162 & 20.24 & 47.1283 & 22.622 & 14.6026 & 29.188 & 0.10 \\
 & 163 & 20.25 & 47.3230 & 22.147 & 14.7975 & 28.728 & 0.35 \\
 & 164 & 20.51 & 46.8612 & 20.236 & 14.3363 & 26.783 & 0.22 \\
 & 165 & 20.60 & 47.1461 & 22.794 & 14.6203 & 29.362 & 0.09 \\
         & & & & & & \\
 & 166 & 20.60 & 46.6794 & 19.235 & 14.1548 & 25.769 & 0.11 \\
 & 167 & 20.60 & 47.1822 & 23.165 & 14.6563 & 29.735 & 0.10 \\
 & 168 & 20.60 & 47.3399 & 21.344 & 14.8146 & 27.926 & 0.12 \\
 & 169 & 20.81 & 47.2581 & 22.024 & 14.7326 & 28.600 & 0.52 \\
 & 170 & 20.95 & 47.3357 & 20.783 & 14.8106 & 27.364 & 0.12 \\
         & & & & & & \\
 & 171 & 20.95 & 47.2488 & 22.793 & 14.7230 & 29.368 & 0.11 \\
 & 172 & 21.10 & 47.2681 & 23.215 & 14.7421 & 29.792 & 0.31 \\
 & 173 & 21.10 & 46.7578 & 18.778 & 14.2334 & 25.318 & 0.12 \\
 & 174 & 21.11 & 46.8273 & 19.902 & 14.3024 & 26.446 & 0.11 \\
 & 175 & 21.19 & 47.2243 & 22.961 & 14.6984 & 29.534 & 0.29 \\
         & & & & & & \\
 & 176 & 21.28 & 47.3091 & 21.994 & 14.7836 & 28.574 & 0.78 \\
 & 177 & 21.30 & 47.0599 & 22.222 & 14.5343 & 28.784 & 0.11 \\
 & 178 & 21.30 & 47.0945 & 22.299 & 14.5689 & 28.863 & 0.11 \\
 & 179 & 21.30 & 46.8812 & 20.791 & 14.3561 & 27.340 & 0.14 \\
 & 180 & 21.43 & 47.1741 & 22.606 & 14.6484 & 29.176 & 0.24 \\
         & & & & & & \\
 & 181 & 21.65 & 47.0019 & 21.539 & 14.4766 & 28.096 & 0.10 \\
 & 182 & 21.66 & 47.2872 & 21.697 & 14.7618 & 28.274 & 0.42 \\
 & 183 & 21.98 & 46.8077 & 19.522 & 14.2830 & 26.065 & 0.18 \\
 & 184 & 22.00 & 46.7810 & 19.400 & 14.2564 & 25.941 & 0.11 \\
 & 185 & 22.06 & 47.1652 & 21.292 & 14.6399 & 27.861 & 0.58 \\
         & & & & & & \\
 & 186 & 22.15 & 47.2621 & 21.690 & 14.7366 & 28.266 & 0.23 \\
 & 187 & 22.21 & 46.7510 & 18.649 & 14.2266 & 25.188 & 0.32 \\
 & 188 & 22.35 & 47.1823 & 22.474 & 14.6566 & 29.045 & 0.11 \\
 & 189 & 22.35 & 47.2012 & 22.210 & 14.6756 & 28.782 & 0.10 \\
 & 190 & 22.35 & 47.2252 & 21.153 & 14.6999 & 27.726 & 0.10 \\
         & & & & & & \\
 & 191 & 22.45 & 47.3097 & 21.363 & 14.7844 & 27.942 & 0.22 \\
 & 192 & 22.52 & 46.7363 & 18.380 & 14.2120 & 24.918 & 0.11 \\
 & 193 & 22.52 & 47.1626 & 21.569 & 14.6372 & 28.137 & 1.16 \\
 & 194 & 22.68 & 46.7764 & 19.147 & 14.2518 & 25.687 & 0.20 \\
 & 195 & 22.70 & 47.2990 & 22.012 & 14.7734 & 28.591 & 0.10 \\
         & & & & & & \\
 & 196 & 22.70 & 47.3204 & 21.683 & 14.7950 & 28.263 & 0.12 \\
 & 197 & 22.89 & 47.2768 & 21.329 & 14.7515 & 27.906 & 0.24 \\
\hline
\end{tabular}
\end{table*}

\begin{table*} 
\addtocounter{table}{-1}
\caption{OH 1.6-GHz Masers in Orion-BN/KL, continued.}
\begin{tabular}{cccccccc}
\hline
Transition &  Feature &  Velocity   &  RA(1950)   &  Dec(1950) &  RA(2000) & Dec(2000) & Intensity  \\ 
           &  No.     &  (km~s$^{-1}$) &  05$^{\rm h}$32$^{\rm m}$  $^{\rm s}$ & --5$^{\circ}$:24\arcmin:  \arcsec &  05$^{\rm h}$h35$^{\rm m}$ $^{\rm s}$  &  --5$^{\circ}$:22\arcmin:  \arcsec & (Jy~beam$^{-1}$) \\
\hline 

{\em 1665L} 

 & 198 & 23.45 & 47.3606 & 19.136 & 14.8360 & 25.719 & 0.14 \\
 & 199 & 23.76 & 47.0703 & 21.506 & 14.5449 & 28.068 & 0.09 \\
 & 200 & 25.52 & 47.1255 & 20.913 & 14.6003 & 27.479 & 0.12 \\
         & & & & & & \\
 & 201 & 30.27 & 47.0935 & 20.382 & 14.5685 & 26.946 & 0.27 \\
 & 202 & 37.12 & 46.9894 & 20.319 & 14.4644 & 26.876 & 0.11 \\
 & 203 & 38.34 & 46.9856 & 20.301 & 14.4607 & 26.857 & 0.17 \\
 & 204 & 38.53 & 47.5898 & 14.835 & 15.0666 & 21.435 & 0.10 \\

\hline
{\em 1665R} 
 & 1 & --10.64 & 46.7892 & 22.184 & 14.2636 & 28.725 & 0.15 \\
 & 2 &  --8.94 & 46.7760 & 21.434 & 14.2507 & 27.975 & 0.13 \\
 & 3 &  --8.59 & 46.8107 & 23.575 & 14.2847 & 30.118 & 0.25 \\
 & 4 &  --8.59 & 46.8138 & 21.872 & 14.2884 & 28.415 & 0.11 \\
 & 5 &  --8.59 & 46.8852 & 19.533 & 14.3605 & 26.082 & 0.10 \\
         & & & & & & \\
 & 6 &  --8.08 & 46.4489 & 22.618 & 13.9232 & 29.135 & 0.19 \\
 & 7 &  --8.04 & 46.9381 & 20.614 & 14.4131 & 27.167 & 0.16 \\
 & 8 &  --7.88 & 46.6323 & 22.058 & 14.1068 & 28.589 & 0.12 \\
 & 9 &  --7.88 & 46.9518 & 20.971 & 14.4266 & 27.525 & 0.12 \\
 & 10 & --7.53 & 46.8734 & 24.560 & 14.3471 & 31.108 & 0.16 \\
         & & & & & & \\
 & 11 & --7.53 & 46.8974 & 20.259 & 14.3724 & 26.809 & 0.13 \\
 & 12 & --7.27 & 46.8497 & 24.540 & 14.3234 & 31.086 & 1.72 \\
 & 13 & --7.00 & 46.9321 & 20.761 & 14.4070 & 27.313 & 0.72 \\
 & 14 & --6.83 & 46.7781 & 21.029 & 14.2529 & 27.570 & 0.12 \\
 & 15 & --6.55 & 46.6520 & 22.005 & 14.1265 & 28.537 & 0.26 \\
         & & & & & & \\
 & 16 & --6.48 & 46.7726 & 21.349 & 14.2473 & 27.889 & 0.20 \\
 & 17 & --6.48 & 46.7973 & 21.298 & 14.2720 & 27.841 & 0.10 \\
 & 18 & --6.23 & 46.6608 & 23.428 & 14.1349 & 29.960 & 1.18 \\
 & 19 & --5.77 & 46.8957 & 23.757 & 14.3696 & 30.307 & 0.24 \\
 & 20 & --5.77 & 46.7760 & 21.290 & 14.2507 & 27.831 & 0.11 \\
         & & & & & & \\
 & 21 & --5.67 & 46.9149 & 23.815 & 14.3888 & 30.365 & 0.48 \\
 & 22 & --5.42 & 46.5014 & 11.572 & 13.9793 & 18.093 & 0.09 \\
 & 23 & --4.25 & 47.0269 & 24.013 & 14.5007 & 30.572 & 0.23 \\
 & 24 & --3.89 & 46.6230 & 21.806 & 14.0976 & 28.336 & 0.18 \\
 & 25 & --3.52 & 46.7739 & 20.874 & 14.2488 & 27.414 & 0.42 \\
         & & & & & & \\
 & 26 & --3.31 & 46.5598 & 21.664 & 14.0344 & 28.189 & 0.26 \\
 & 27 & --3.31 & 46.4710 & 21.594 & 13.9456 & 28.113 & 0.11 \\
 & 28 & --3.31 & 47.6469 & 23.613 & 15.1208 & 30.217 & 0.10 \\
 & 29 & --1.91 & 46.8620 & 20.278 & 14.3370 & 26.825 & 0.20 \\
 & 30 & --1.35 & 46.9597 & 23.737 & 14.4336 & 30.291 & 0.93 \\
         & & & & & & \\
 & 31 & --0.50 & 47.5735 & 25.373 & 15.0469 & 31.971 & 0.12 \\
 & 32 &  0.02 & 47.1037 & 21.252 & 14.5785 & 27.816 & 1.82 \\
 & 33 &  0.55 & 46.8028 & 23.592 & 14.2767 & 30.135 & 0.15 \\
 & 34 &  0.72 & 46.9623 & 23.565 & 14.4363 & 30.119 & 3.03 \\
 & 35 &  1.15 & 46.7232 & 20.962 & 14.1980 & 27.499 & 0.30 \\
         & & & & & & \\
 & 36 &  1.87 & 46.9477 & 23.778 & 14.4216 & 30.331 & 4.29 \\
 & 37 &  1.96 & 46.9203 & 23.109 & 14.3944 & 29.660 & 0.10 \\
 & 38 &  1.96 & 46.9669 & 24.141 & 14.4407 & 30.696 & 0.10 \\
 & 39 &  2.31 & 46.7483 & 21.446 & 14.2230 & 27.985 & 0.10 \\
 & 40 &  2.34 & 46.6527 & 20.863 & 14.1275 & 27.395 & 0.28 \\
         & & & & & & \\
 & 41 &  2.43 & 46.7296 & 20.860 & 14.2044 & 27.397 & 0.27 \\
 & 42 &  2.43 & 46.7989 & 23.565 & 14.2729 & 30.108 & 1.48 \\
 & 43 &  2.44 & 46.7811 & 21.544 & 14.2557 & 28.085 & 0.24 \\
 & 44 &  2.90 & 46.5691 & 21.459 & 14.0438 & 27.984 & 0.19 \\
 & 45 &  3.02 & 46.9402 & 22.635 & 14.4144 & 29.188 & 0.29 \\
\hline
\end{tabular}
\end{table*}

\begin{table*} 
\addtocounter{table}{-1}
\caption{OH 1.6-GHz Masers in Orion-BN/KL, continued.}
\begin{tabular}{cccccccc}
\hline
Transition &  Feature &  Velocity   &  RA(1950)   &  Dec(1950) &  RA(2000) & Dec(2000) & Intensity  \\ 
           &  No.     &  (km~s$^{-1}$) &  05$^{\rm h}$32$^{\rm m}$  $^{\rm s}$ & --5$^{\circ}$:24\arcmin:  \arcsec &  05$^{\rm h}$h35$^{\rm m}$ $^{\rm s}$  &  --5$^{\circ}$:22\arcmin:  \arcsec & (Jy~beam$^{-1}$) \\
\hline 

{\em 1665R} 
 & 46 &  3.53 & 46.7474 & 20.716 & 14.2223 & 27.254 & 1.01 \\
 & 47 &  3.61 & 46.6689 & 20.895 & 14.1438 & 27.428 & 0.39 \\
 & 48 &  3.72 & 46.7300 & 20.675 & 14.2049 & 27.213 & 0.23 \\
 & 49 &  4.07 & 46.5933 & 21.349 & 14.0680 & 27.877 & 0.78 \\
 & 50 &  4.07 & 46.5446 & 21.273 & 14.0194 & 27.797 & 0.21 \\ 
         & & & & & & \\
 & 51 &  4.24 & 46.8052 & 23.576 & 14.2792 & 30.119 & 0.18 \\
 & 52 &  4.42 & 46.7489 & 20.825 & 14.2238 & 27.364 & 0.21 \\
 & 53 &  4.77 & 46.6189 & 21.236 & 14.0937 & 27.765 & 0.19 \\
 & 54 &  5.03 & 46.7662 & 20.640 & 14.2412 & 27.180 & 1.01 \\
 & 55 &  5.13 & 47.0036 & 23.137 & 14.4777 & 29.694 & 0.13 \\
         & & & & & & \\
 & 56 &  5.13 & 47.0114 & 21.280 & 14.4861 & 27.837 & 0.10 \\
 & 57 &  5.42 & 46.9331 & 24.058 & 14.4069 & 30.610 & 13.32 \\
 & 58 &  5.48 & 46.9861 & 22.727 & 14.4603 & 29.283 & 0.13 \\
 & 59 &  5.48 & 46.9150 & 22.140 & 14.3895 & 28.691 & 0.11 \\
 & 60 &  5.48 & 46.9950 & 21.363 & 14.4697 & 27.920 & 0.12 \\
         & & & & & & \\
 & 61 &  5.48 & 46.9345 & 23.217 & 14.4086 & 29.769 & 0.11 \\
 & 62 &  5.60 & 46.9629 & 22.917 & 14.4371 & 29.472 & 1.97 \\
 & 63 &  5.63 & 47.1711 & 34.641 & 14.6414 & 41.210 & 0.13 \\
 & 64 &  5.66 & 47.1858 & 34.679 & 14.6562 & 41.249 & 0.15 \\
 & 65 &  5.83 & 46.6880 & 20.712 & 14.1630 & 27.246 & 0.12 \\
         & & & & & & \\
 & 66 &  5.83 & 46.3557 & 19.353 & 13.8311 & 25.863 & 0.12 \\
 & 67 &  6.18 & 46.6603 & 20.783 & 14.1352 & 27.316 & 0.12 \\
 & 68 &  6.18 & 46.4246 & 18.394 & 13.9003 & 24.909 & 0.12 \\
 & 69 &  6.18 & 46.8949 & 20.286 & 14.3700 & 26.836 & 0.29 \\
 & 70 &  6.29 & 46.3697 & 18.686 & 13.8453 & 25.197 & 0.30 \\
         & & & & & & \\
 & 71 &  6.53 & 46.9588 & 22.910 & 14.4330 & 29.464 & 0.14 \\
 & 72 &  6.61 & 46.7039 & 20.616 & 14.1789 & 27.151 & 1.09 \\
 & 73 &  6.63 & 46.4055 & 18.617 & 13.8811 & 25.131 & 0.52 \\
 & 74 &  6.88 & 47.4773 & 19.497 & 14.9526 & 26.089 & 0.10 \\
 & 75 &  6.88 & 47.4471 & 19.446 & 14.9224 & 26.035 & 0.11 \\
         & & & & & & \\
 & 76 &  7.01 & 47.2481 & 35.460 & 14.7182 & 42.034 & 0.20 \\
 & 77 &  7.11 & 46.6565 & 20.807 & 14.1314 & 27.339 & 0.49 \\
 & 78 &  7.12 & 46.7840 & 24.318 & 14.2577 & 30.860 & 0.32 \\
 & 79 &  7.23 & 46.9036 & 24.377 & 14.3773 & 30.927 & 3.04 \\
 & 80 &  7.43 & 47.4634 & 19.557 & 14.9387 & 26.148 & 0.55 \\
         & & & & & & \\
 & 81 &  7.59 & 46.5732 & 20.572 & 14.0482 & 27.098 & 0.10 \\
 & 82 &  7.59 & 46.7991 & 24.012 & 14.2729 & 30.554 & 0.11 \\
 & 83 &  7.59 & 46.5521 & 21.000 & 14.0269 & 27.525 & 0.14 \\
 & 84 &  7.73 & 46.4607 & 35.073 & 13.9310 & 41.591 & 0.88 \\
 & 85 &  7.94 & 46.6746 & 20.700 & 14.1495 & 27.233 & 0.10 \\
         & & & & & & \\
 & 86 &  7.94 & 46.6647 & 29.883 & 14.1367 & 36.416 & 0.11 \\
 & 87 &  8.63 & 47.4493 & 19.666 & 14.9245 & 26.255 & 0.22 \\
 & 88 &  8.64 & 46.6916 & 29.404 & 14.1637 & 35.938 & 0.28 \\
 & 89 &  8.64 & 46.8272 & 23.964 & 14.3011 & 30.508 & 0.10 \\
 & 90 &  8.64 & 47.1435 & 21.659 & 14.6181 & 28.227 & 0.15 \\
         & & & & & & \\
 & 91 &  8.79 & 47.3896 & 18.152 & 14.8653 & 24.737 & 0.15 \\
 & 92 &  8.84 & 46.8000 & 24.518 & 14.2737 & 31.060 & 0.17 \\
 & 93 &  8.99 & 47.6212 & 20.111 & 15.0962 & 26.713 & 0.11 \\
 & 94 &  8.99 & 47.0118 & 23.011 & 14.4860 & 29.569 & 0.14 \\
 & 95 &  8.99 & 46.6573 & 20.772 & 14.1322 & 27.304 & 0.12 \\
         & & & & & & \\
 & 96 &  9.18 & 47.5979 & 19.677 & 15.0731 & 26.277 & 0.22 \\
 & 97 &  9.34 & 47.5998 & 19.994 & 15.0749 & 26.594 & 0.10 \\
\hline
\end{tabular}
\end{table*}

\begin{table*} 
\addtocounter{table}{-1}
\caption{OH 1.6-GHz Masers in Orion-BN/KL, continued.}
\begin{tabular}{cccccccc}
\hline
Transition &  Feature &  Velocity   &  RA(1950)   &  Dec(1950) &  RA(2000) & Dec(2000) & Intensity  \\ 
           &  No.     &  (km~s$^{-1}$) &  05$^{\rm h}$32$^{\rm m}$  $^{\rm s}$ & --5$^{\circ}$:24\arcmin:  \arcsec &  05$^{\rm h}$h35$^{\rm m}$ $^{\rm s}$  &  --5$^{\circ}$:22\arcmin:  \arcsec & (Jy~beam$^{-1}$) \\
\hline 

{\em 1665R} 

 & 98 &  9.34 & 47.0384 & 22.393 & 14.5127 & 28.953 & 0.12 \\
 & 99 &  9.34 & 47.5797 & 19.486 & 15.0550 & 26.085 & 0.13 \\
 & 100 &  9.49 & 46.8978 & 20.275 & 14.3729 & 26.825 & 0.14 \\
         & & & & & & \\
 & 101 & 10.05 & 46.9781 & 22.586 & 14.4524 & 29.141 & 0.24 \\
 & 102 & 10.70 & 46.7569 & 31.179 & 14.2284 & 37.718 & 0.34 \\
 & 103 & 10.75 & 47.3695 & 21.663 & 14.8441 & 28.247 & 0.14 \\
 & 104 & 10.79 & 47.0691 & 21.578 & 14.5437 & 28.140 & 1.26 \\
 & 105 & 11.10 & 46.6243 & 20.873 & 14.0992 & 27.403 & 0.24 \\
         & & & & & & \\
 & 106 & 11.10 & 47.4608 & 18.220 & 14.9365 & 24.810 & 0.12 \\
 & 107 & 11.25 & 47.3932 & 18.070 & 14.8689 & 24.656 & 0.45 \\
 & 108 & 11.81 & 47.3321 & 10.317 & 14.8104 & 16.898 & 0.12 \\
 & 109 & 11.81 & 47.4134 & 23.297 & 14.8874 & 29.884 & 0.12 \\
 & 110 & 11.81 & 47.4272 & 21.925 & 14.9017 & 28.512 & 1.46 \\
         & & & & & & \\
 & 111 & 12.16 & 47.0128 & 21.766 & 14.4874 & 28.324 & 0.19 \\
 & 112 & 12.16 & 46.9665 & 21.188 & 14.4412 & 27.743 & 0.14 \\
 & 113 & 12.16 & 46.6537 & 20.600 & 14.1287 & 27.132 & 0.15 \\
 & 114 & 12.24 & 47.4024 & 21.939 & 14.8769 & 28.525 & 0.39 \\
 & 115 & 12.51 & 47.2698 & 17.198 & 14.7458 & 23.775 & 0.13 \\
         & & & & & & \\
 & 116 & 12.72 & 47.0583 & 22.027 & 14.5328 & 28.589 & 0.19 \\
 & 117 & 12.86 & 47.0778 & 21.726 & 14.5523 & 28.288 & 0.14 \\
 & 118 & 13.07 & 46.8463 & 20.094 & 14.3215 & 26.640 & 0.17 \\
 & 119 & 13.56 & 47.4364 & 18.336 & 14.9120 & 24.924 & 0.17 \\
 & 120 & 13.56 & 46.7487 & 19.915 & 14.2239 & 26.454 & 0.22 \\
         & & & & & & \\
 & 121 & 13.56 & 47.0897 & 21.697 & 14.5643 & 28.260 & 0.13 \\
 & 122 & 13.99 & 47.0696 & 22.149 & 14.5440 & 28.711 & 0.36 \\
 & 123 & 14.27 & 46.8284 & 10.361 & 14.3066 & 16.906 & 0.10 \\
 & 124 & 14.39 & 47.0897 & 22.426 & 14.5641 & 28.989 & 0.35 \\
 & 125 & 14.76 & 46.7700 & 19.827 & 14.2452 & 26.367 & 0.28 \\
         & & & & & & \\
 & 126 & 15.67 & 47.0976 & 22.393 & 14.5720 & 28.957 & 0.15 \\
 & 127 & 15.67 & 46.8512 & 19.455 & 14.3265 & 26.001 & 0.12 \\
 & 128 & 15.67 & 46.7914 & 19.479 & 14.2667 & 26.021 & 0.12 \\
 & 129 & 15.67 & 47.1407 & 22.790 & 14.6149 & 29.357 & 0.09 \\
 & 130 & 15.67 & 47.1195 & 22.518 & 14.5938 & 29.083 & 0.10 \\
         & & & & & & \\
 & 131 & 15.83 & 47.3816 & 21.585 & 14.8562 & 28.169 & 0.49 \\
 & 132 & 16.38 & 47.1348 & 22.297 & 14.6092 & 28.864 & 0.16 \\
 & 133 & 16.38 & 47.0745 & 21.293 & 14.5492 & 27.855 & 0.11 \\
 & 134 & 16.73 & 47.1427 & 21.852 & 14.6172 & 28.419 & 0.14 \\
 & 135 & 16.73 & 47.4460 & 21.058 & 14.9208 & 27.647 & 0.10 \\
         & & & & & & \\
 & 136 & 16.73 & 47.1279 & 22.473 & 14.6022 & 29.039 & 0.14 \\
 & 137 & 17.08 & 47.1074 & 22.284 & 14.5818 & 28.849 & 0.12 \\
 & 138 & 17.30 & 47.1830 & 21.759 & 14.6576 & 28.329 & 0.49 \\
 & 139 & 17.43 & 46.7458 & 19.202 & 14.2212 & 25.740 & 0.10 \\
 & 140 & 17.43 & 47.2621 & 22.741 & 14.7363 & 29.317 & 0.11 \\
         & & & & & & \\
 & 141 & 17.52 & 47.2124 & 22.805 & 14.6866 & 29.377 & 0.33 \\
 & 142 & 17.70 & 47.1094 & 22.098 & 14.5839 & 28.662 & 1.33 \\
 & 143 & 17.76 & 47.2777 & 23.112 & 14.7518 & 29.689 & 0.16 \\
 & 144 & 17.85 & 47.1364 & 22.364 & 14.6108 & 28.931 & 1.05 \\
 & 145 & 17.89 & 47.0646 & 22.633 & 14.5389 & 29.194 & 0.26 \\
         & & & & & & \\
 & 146 & 17.99 & 46.6726 & 20.281 & 14.1477 & 26.814 & 0.18 \\
 & 147 & 18.49 & 47.1717 & 22.117 & 14.6462 & 28.687 & 0.11 \\
 & 148 & 18.49 & 47.1378 & 24.08s & 14.6116 & 30.649 & 0.11 \\
 & 149 & 18.49 & 46.9396 & 11.550 & 14.4174 & 18.103 & 0.10 \\

\hline
\end{tabular}
\end{table*}

\begin{table*} 
\addtocounter{table}{-1}
\caption{OH 1.6-GHz Masers in Orion-BN/KL, continued.}
\begin{tabular}{cccccccc}
\hline
Transition &  Feature &  Velocity   &  RA(1950)   &  Dec(1950) &  RA(2000) & Dec(2000) & Intensity  \\ 
           &  No.     &  (km~s$^{-1}$) &  05$^{\rm h}$32$^{\rm m}$  $^{\rm s}$ & --5$^{\circ}$:24\arcmin:  \arcsec &  05$^{\rm h}$h35$^{\rm m}$ $^{\rm s}$  &  --5$^{\circ}$:22\arcmin:  \arcsec & (Jy~beam$^{-1}$) \\
\hline 

{\em 1665R} 
 & 150 & 18.49 & 47.2269 & 22.836 & 14.7011 & 29.409 & 0.12 \\
         & & & & & & \\
 & 151 & 18.49 & 47.1035 & 23.022 & 14.5776 & 29.586 & 0.10 \\
 & 152 & 18.49 & 47.1097 & 23.543 & 14.5836 & 30.108 & 0.08 \\
 & 153 & 18.49 & 47.3616 & 21.744 & 14.8361 & 28.327 & 0.12 \\
 & 154 & 18.49 & 47.0872 & 24.091 & 14.5610 & 30.655 & 0.08 \\
 & 155 & 18.63 & 47.1870 & 23.139 & 14.6611 & 29.709 & 0.23 \\
         & & & & & & \\
 & 156 & 18.64 & 47.1385 & 22.269 & 14.6128 & 28.836 & 7.34 \\
 & 157 & 18.76 & 47.1659 & 22.942 & 14.6401 & 29.511 & 0.77 \\
 & 158 & 18.84 & 47.1304 & 21.627 & 14.6050 & 28.194 & 0.10 \\
 & 159 & 18.84 & 46.6699 & 20.266 & 14.1449 & 26.799 & 0.12 \\
 & 160 & 18.84 & 47.1244 & 23.040 & 14.5986 & 29.606 & 0.15 \\
         & & & & & & \\
 & 161 & 18.84 & 47.1112 & 22.301 & 14.5856 & 28.867 & 0.12 \\
 & 162 & 18.84 & 47.1303 & 23.436 & 14.6043 & 30.002 & 0.09 \\
 & 163 & 18.84 & 47.0534 & 23.996 & 14.5273 & 30.557 & 0.10 \\
 & 164 & 19.03 & 46.7193 & 19.180 & 14.1947 & 25.716 & 0.18 \\
 & 165 & 19.19 & 47.3167 & 22.621 & 14.7910 & 29.201 & 0.11 \\
         & & & & & & \\
 & 166 & 19.19 & 47.0641 & 22.715 & 14.5383 & 29.276 & 0.09 \\
 & 167 & 19.35 & 46.9618 & 21.237 & 14.4365 & 27.792 & 0.11 \\
 & 168 & 19.42 & 47.3386 & 22.245 & 14.8129 & 28.826 & 0.21 \\
 & 169 & 19.46 & 47.1366 & 22.211 & 14.6110 & 28.778 & 0.37 \\
 & 170 & 19.54 & 47.1549 & 22.679 & 14.6292 & 29.247 & 0.16 \\
         & & & & & & \\
 & 171 & 19.54 & 47.2993 & 23.040 & 14.7734 & 29.618 & 0.10 \\
 & 172 & 19.54 & 47.2746 & 22.420 & 14.7489 & 28.997 & 0.25 \\
 & 173 & 19.54 & 46.6645 & 19.663 & 14.1398 & 26.196 & 0.11 \\
 & 174 & 19.54 & 47.0678 & 22.465 & 14.5422 & 29.027 & 0.16 \\
 & 175 & 19.54 & 47.3164 & 22.148 & 14.7909 & 28.728 & 0.09 \\
         & & & & & & \\
 & 176 & 19.54 & 47.1038 & 22.720 & 14.5780 & 29.284 & 0.16 \\
 & 177 & 19.71 & 47.1870 & 23.004 & 14.6612 & 29.575 & 0.39 \\
 & 178 & 19.89 & 47.2796 & 21.609 & 14.7542 & 28.187 & 0.10 \\
 & 179 & 19.89 & 47.3264 & 22.292 & 14.8008 & 28.872 & 0.11 \\
 & 180 & 20.00 & 47.2892 & 22.323 & 14.7635 & 28.901 & 4.24 \\
         & & & & & & \\
 & 181 & 20.06 & 47.2165 & 23.524 & 14.6905 & 30.097 & 0.18 \\
 & 182 & 20.09 & 46.9299 & 14.804 & 14.4068 & 21.356 & 0.12 \\
 & 183 & 20.24 & 47.1416 & 22.404 & 14.6159 & 28.972 & 0.21 \\
 & 184 & 20.24 & 47.3210 & 22.020 & 14.7954 & 28.600 & 0.38 \\
 & 185 & 20.24 & 47.3613 & 21.756 & 14.8359 & 28.339 & 0.10 \\
         & & & & & & \\
 & 186 & 20.24 & 46.8675 & 20.300 & 14.3425 & 26.848 & 0.11 \\
 & 187 & 20.24 & 47.2083 & 22.882 & 14.6825 & 29.454 & 0.20 \\
 & 188 & 20.24 & 47.2757 & 22.359 & 14.7500 & 28.936 & 0.23 \\
 & 189 & 20.24 & 47.2389 & 23.568 & 14.7129 & 30.142 & 0.14 \\
 & 190 & 20.49 & 46.8274 & 19.833 & 14.3026 & 26.378 & 0.34 \\
         & & & & & & \\
 & 191 & 20.60 & 47.2254 & 22.594 & 14.6997 & 29.167 & 0.28 \\
 & 192 & 20.61 & 47.1927 & 22.700 & 14.6670 & 29.271 & 0.28 \\
 & 193 & 20.73 & 47.0252 & 18.005 & 14.5010 & 24.563 & 0.17 \\
 & 194 & 20.78 & 47.2590 & 22.030 & 14.7335 & 28.605 & 0.59 \\
 & 195 & 20.79 & 47.1323 & 22.381 & 14.6066 & 28.947 & 0.32 \\
         & & & & & & \\
 & 196 & 20.95 & 47.1747 & 22.706 & 14.6489 & 29.276 & 0.12 \\
 & 197 & 21.11 & 47.3075 & 21.887 & 14.7820 & 28.466 & 0.72 \\
 & 198 & 21.30 & 47.2234 & 23.040 & 14.6976 & 29.613 & 0.10 \\
 & 199 & 21.30 & 47.2034 & 23.131 & 14.6775 & 29.703 & 0.10 \\
 & 200 & 21.30 & 47.1751 & 22.577 & 14.6494 & 29.147 & 0.31 \\
         & & & & & & \\
\hline
\end{tabular}
\end{table*}

\begin{table*} 
\addtocounter{table}{-1}
\caption{OH 1.6-GHz Masers in Orion-BN/KL, continued.}
\begin{tabular}{cccccccc}
\hline
Transition &  Feature &  Velocity   &  RA(1950)   &  Dec(1950) &  RA(2000) & Dec(2000) & Intensity  \\ 
           &  No.     &  (km~s$^{-1}$) &  05$^{\rm h}$32$^{\rm m}$  $^{\rm s}$ & --5$^{\circ}$:24\arcmin:  \arcsec &  05$^{\rm h}$h35$^{\rm m}$ $^{\rm s}$  &  --5$^{\circ}$:22\arcmin:  \arcsec & (Jy~beam$^{-1}$) \\
\hline 

{\em 1665R} 
 & 201 & 21.38 & 47.2850 & 21.772 & 14.7596 & 28.350 & 1.48 \\
 & 202 & 21.51 & 46.7892 & 19.337 & 14.2645 & 25.879 & 0.18 \\
 & 203 & 21.65 & 46.7448 & 18.535 & 14.2205 & 25.074 & 0.14 \\
 & 204 & 22.00 & 47.1648 & 22.167 & 14.6392 & 28.736 & 0.15 \\
 & 205 & 22.00 & 47.2413 & 21.959 & 14.7158 & 28.534 & 0.15 \\
         & & & & & & \\
 & 206 & 22.00 & 47.1936 & 22.864 & 14.6678 & 29.435 & 0.12 \\
 & 207 & 22.00 & 47.1771 & 21.131 & 14.6518 & 27.701 & 0.22 \\
 & 208 & 22.01 & 46.7630 & 18.902 & 14.2385 & 25.442 & 0.19 \\
 & 209 & 22.22 & 47.3115 & 21.333 & 14.7862 & 27.913 & 0.29 \\
 & 210 & 22.35 & 46.7742 & 19.028 & 14.2497 & 25.569 & 0.11 \\
         & & & & & & \\
 & 211 & 22.35 & 46.8020 & 19.538 & 14.2773 & 26.081 & 0.08 \\
 & 212 & 22.50 & 47.1623 & 21.593 & 14.6369 & 28.162 & 2.19 \\
 & 213 & 22.54 & 46.6625 & 23.668 & 14.1364 & 30.201 & 0.84 \\
 & 214 & 22.58 & 47.2731 & 21.427 & 14.7478 & 28.004 & 0.20 \\
 & 215 & 22.70 & 47.2730 & 20.298 & 14.7480 & 26.875 & 0.09 \\
         & & & & & & \\
 & 216 & 22.70 & 46.7332 & 18.448 & 14.2089 & 24.986 & 0.11 \\
 & 217 & 22.70 & 46.7309 & 20.211 & 14.2060 & 26.748 & 0.10 \\
 & 218 & 22.70 & 46.8040 & 19.311 & 14.2794 & 25.854 & 0.09 \\
 & 219 & 22.70 & 47.2795 & 19.234 & 14.7549 & 25.811 & 0.31 \\
 & 220 & 22.81 & 47.3016 & 21.029 & 14.7764 & 27.608 & 0.25 \\
         & & & & & & \\
 & 221 & 22.86 & 47.3174 & 21.643 & 14.7920 & 28.223 & 0.26 \\
 & 222 & 23.06 & 46.8216 & 19.733 & 14.2968 & 26.277 & 0.13 \\
 & 223 & 23.06 & 47.0926 & 21.834 & 14.5671 & 28.397 & 0.09 \\
 & 224 & 23.06 & 46.8102 & 19.504 & 14.2855 & 26.047 & 0.10 \\
 & 225 & 23.57 & 47.3143 & 19.297 & 14.7897 & 25.877 & 0.10 \\
         & & & & & & \\
 & 226 & 23.65 & 47.3609 & 19.201 & 14.8363 & 25.784 & 0.31 \\

\hline
{\em 1667L}  & 1    & 13.46 &  46.9359 &  24.607 & 14.4096 & 31.159 & 0.30  \\ 
             & 2    & 13.51 &  46.9845 &  24.415 & 14.4582 & 30.971 & 0.21  \\
             & 3    & 13.43 &  46.9832 &  29.620 & 14.4552 & 36.176 & .21  \\
\hline
\end{tabular}
\end{table*}

\end{document}